\documentclass[preprint2]{aastex631}

\begin{document}

\title{Observing atmospheric escape in sub-Jovian worlds with \emph{JWST}}

\author[0000-0002-2248-3838]{Leonardo A. Dos Santos}
\affiliation{Space Telescope Science Institute, 3700 San Martin Drive, Baltimore, MD 21218, USA}
\email{ldsantos@stsci.edu}

\author[0000-0003-4157-832X]{Munazza K. Alam}
\affiliation{Carnegie Earth \& Planets Laboratory, 5241 Broad Branch Road NW, Washington, DC 20015, USA}

\author[0000-0001-9513-1449]{N\'estor Espinoza}
\affiliation{Space Telescope Science Institute, 3700 San Martin Drive, Baltimore, MD 21218, USA}

\author[0000-0003-2527-1475]{Shreyas~Vissapragada}
\altaffiliation{51 Pegasi b Fellow}
\affil{Center for Astrophysics $\vert$ Harvard \& Smithsonian, 60 Garden Street, Cambridge, MA 02138, USA}


\begin{abstract}

Hydrodynamic atmospheric escape is considered an important process that shapes the evolution of sub-Jovian exoplanets, particularly those with short orbital periods. The metastable He line in the near-infrared at $1.083$~$\mu$m is a reliable tracer of atmospheric escape in hot exoplanets, with the advantage of being observable from the ground. However, observing escaping He in sub-Jovian planets has remained challenging due to the systematic effects and telluric contamination present in ground-based data. With the successful launch and operations of \emph{JWST}, we now have access to extremely stable high-precision near-infrared spectrographs in space. Here we predict the observability of metastable He with \emph{JWST} in two representative and previously well-studied warm Neptunes, GJ~436~b ($T_{\rm eq} = 687~{\rm K}$, $R_{\rm p} = 0.37~{\rm R_J}$) and GJ~1214~b ($T_{\rm eq} = 588~{\rm K}$, $R_{\rm p} = 0.25~{\rm R_J}$). Our simulated \emph{JWST} observations for GJ~436~b demonstrate that a single transit with NIRSpec/G140H is sensitive to mass loss rates that are two orders of magnitude lower than what is detectable from the ground. Our exercise for GJ~1214~b show that the best configuration to observe the relatively weak outflows of warm Neptunes with \emph{JWST} is with NIRSpec/G140H, and that NIRSpec/G140M and NIRISS/SOSS are less optimal. Since none of these instrument configurations can spectrally resolve the planetary absorption, we conclude that the 1D isothermal Parker-wind approximation may not be sufficient for interpreting such observations. More sophisticated models are critical for breaking the degeneracy between outflow temperature and mass-loss rate for \emph{JWST} measurements of metastable He. 

\end{abstract}



\section{Introduction} \label{sec:intro}

Surveys for transiting exoplanets have revealed two demographic features that are linked to the evolution of exoplanets: the hot Neptune desert \citep{Szabo2011, Beauge2013, Mazeh2016} and the radius valley \citep{Fulton2017, Fulton2018}. Short-period, sub-Jovian exoplanets may rapidly change in size in their early lives due to atmospheric escape driven by a combination of intense high-energy irradiation from their host star \citep[e.g.,][]{Lammer2003, Chadney2015, Owen2013, Owen2017, Ionov2018, Mordasini2020} and their internal energy residual from formation \citep{Ginzburg2018, Gupta2019}. Prior to this discovery, hydrodynamic atmospheric escape (or evaporation) had already been predicted for the early Solar System planets \citep{Chamberlain1963, Watson1981, Hunten1987}, observed in hot Jupiters \citep{Vidal2003, Vidal2004, Lecavelier2010, Fossati2010}, and later in Neptunes as well \citep{Ehrenreich2015, Bourrier2018}.

Classically, these observations were first carried out with the ultraviolet (UV) capabilities of the \emph{Hubble Space Telescope} (\emph{HST}). In these wavelengths, we have access to escaping neutral H through the Lyman-$\alpha$ line (1216.67~\AA), as well as metallic species like C, N, O, Si, Mg and Fe \citep[e.g.,][]{Linsky2010, Vidal2013, Ballester2015, Sing2019, DSantos2019, Garcia2021}. Some of the challenges of these UV observations arise from the relatively lower efficiency of UV detectors compared to optical or infrared (IR) instruments, the lower observable fluxes of cool stars in the UV, interstellar medium absorption, and the fact that \emph{HST} is the only telescope capable of UV spectroscopy currently available to the community.

With the discovery of metastable He as a reliable tracer for atmospheric escape in the near-IR \citep{Oklopcic2018, Allart2018, Spake2018, Mansfield2018}, transit observations at 1.083~$\mu$m have been particularly productive and quickly surpassed the number of evaporation detections compared to UV campaigns \citep[see, e.g.,][]{Kirk2022, Orell2022, DSantos2023}. The advantage of these observations is clear: accessibility from the ground permits more systematic surveys using narrow-band photometry \citep[e.g.,][]{Vissapragada2022b}, as well as the opportunity to resolve planetary absorption at high spectral resolution. Some of the disadvantages of this technique, however, include the fact that populating the metastable He state requires relatively high levels of extreme-ultraviolet (EUV) flux compared to mid-ultraviolet (mid-UV) flux, a condition usually satisfied for K-type stars \citep{Oklopcic2019}. \citet{Poppenhaeger2022} also suggested that stellar coronal abundances have an important role in this process.

Non-detections of metastable He in planets that are expected to be evaporating have been a curious and yet unexplored outcome of observational surveys to date \citep[see, e.g., the case of WASP-80~b;][]{Fossati2022, Vissapragada2022b}. In particular, the warm Neptune GJ~436~b has been found to be enshrouded in a large cloud of neutral H fed by atmospheric escape using \emph{HST} \citep{Ehrenreich2015, Lavie2017, DSantos2019}. But He observations with the CARMENES spectrograph yielded a non-detection of escaping He \citep{Nortmann2018}. The inferred atmospheric escape rate for GJ~436~b based on the \emph{HST} observations range from $\sim 10^8$~g\,s$^{-1}$ \citep{Bourrier2016} to $\sim 10^{10}$~g\,s$^{-1}$ \citep{Villarreal2021}. In comparison, the escape rate in hot Jupiters are in the order of $\sim 10^{11}$~g\,s$^{-1}$ \citep[e.g.,][]{Erkaev2015, Salz2016}. Another curious case is that of GJ~1214~b, whose observations with Keck II/NIRSPEC resulted in non-detections \citep{Kasper2020, Spake2022}, but a transit observed with CARMENES yielded a tentative detection \citep{Orell2022}. Other sub-Jovian worlds for which we expected to observe He escape but only non-detections were reported include K2-100~b \citep{Gaidos2020}, AU~Mic~b \citep{Hirano2020}, HD~97658~b \citep{Kasper2020}, K2-136~c \citep{Gaidos2021}, and the systems V1298~Tau \citep{Vissapragada2021}, GJ~9827~b \citep{Kasper2020, Carleo2021}, and HD~63433 \citep{Zhang2022}.

With the successful launch and operations of \emph{JWST}, we now have access to an extremely precise near-IR space telescope that can potentially give us access to observations of escaping He in sub-Jovian worlds. \citet{Fu2022} demonstrated that the NIRISS spectrograph can detect metastable He in the atmosphere of the hot Saturn HAT-P-18~b. This not only confirms the previous detection from ground-based narrow-band photometry \citep{Paragas2021}, but also reveals the presence of a potential He tail trailing the planet \citep[similar to the case of WASP-107~b;][]{Spake2021}. Trailing tails can be missed in narrow-band photometry because the techniques used to remove systematic effects \citep[see, e.g., Section 3 of][]{Paragas2021} can also remove the astrophysical signatures of asymmetric transit light curves.

In this manuscript, we place constraints on the detectability of escaping He in exoplanets with \emph{JWST}. We simulate the outflow of the representative sub-Jovian worlds GJ~436~b and GJ~1214~b, the metastable He radial profile, and their theoretical transmission spectra at infinite resolution. We also constrain the line-spread function of the NIRSpec spectrograph in the G140H and G140M modes, and use this information to predict the observable transmission spectra. We compare the predicted observable signals and assess the constraints on atmospheric escape that these signals can provide. Finally, we assess which instrumental configuration is best for observing metastable He in transiting exoplanets.

This paper has the following structure: in \S \ref{sect:methods}, we describe the theoretical and instrumental setup used in this exercise. In \S \ref{sect:results}, we discuss the results for GJ~436~b and GJ~1214~b, and their interpretation using a one-dimensional Parker-wind model. \S \ref{sect:conclusions} describes our conclusions and recommendations for future observations.

\section{Theoretical and instrumental setup}\label{sect:methods}

There are two instruments available on \emph{JWST} that are capable of observing the near-IR He line: NIRISS and NIRSpec. The first has a spectral resolving power of $R \sim 650$ at 1.083~$\mu$m (wavelength range 0.6-2.8~$\mu$m), while the latter can observe at $R \sim 100$, $R \sim 1000$ and $R \sim 2700$, respectively, for the PRISM (0.5-5.0~$\mu$m), G140M and G140H modes (0.80-1.27~$\mu$m with the F070LP filter; 0.97-1.84~$\mu$m with F100LP). Ideally, higher spectral resolution is desired for He observations because resolving the planetary absorption yields stronger constraints on the outflow temperature -- which in turn produces a better mass loss constraint, since this estimate is degenerate with the outflow temperature. However, even for spectrally unresolved observations, it is still possible to obtain useful constraints for the mass loss rate if we take into account the energetics of the outflow \citep{Vissapragada2022a, Linssen2022}. For the particular case of GJ~436 ($J_{\rm mag} = 6.9$), the only instrument configuration on \emph{JWST} that does not result in saturation is G140H; fainter stars, such as GJ~1214, are observable at lower spectral resolutions.

We simulate the planetary outflows for these two warm Neptunes using the open-source Python framework {\tt p-winds} \citep[][]{DSantos2022}. This code assumes that the outflow can be simplified to an isothermal Parker-wind model \citep{Parker1958} where the mass loss rate, outflow temperature, and H/He number fraction are free parameters \citep{Oklopcic2018, Lampon2020}. We use version 1.3.4\footnote{\footnotesize{The code is available at \url{https://github.com/ladsantos/p-winds}.}}, which includes tidal gravity effects \citep[for more details, see][]{Vissapragada2022b}.

For the first exercise, we aim to assess what mass-loss constraints can readily be obtained from \emph{JWST} observations. We consider the warm Neptune GJ~436~b as a representative case of a planet known to be undergoing atmospheric mass loss at a rate likely too low to be observable from the ground. We assume a sub-stellar escape rate\footnote{\footnotesize{The term sub-stellar is used to refer to the mass-loss rate assuming that the planet is irradiated over 4$\pi$~sr. This assumption is used in one-dimensional models like {\tt p-winds}. In reality, planets are irradiated over $\pi$~sr only, and the total mass-loss rate is obtained by dividing the sub-stellar rate by four.}} of $10^{8}$~g\,s$^{-1}$ and an outflow temperature of $2700$~K, based on the estimates from Lyman-$\alpha$ observations \citep{Bourrier2016, Villarreal2021}. 

Since GJ~436 is observable only with NIRSpec/G140H, the second exercise is to simulate observations of GJ~1214~b to test the performance of the NIRSpec/G140M and NIRISS/SOSS modes in comparison with G140H. Theoretical predictions for this planet estimate a sub-stellar mass-loss rate of $1.9 \times 10^{10}$~g\,s$^{-1}$ and outflow temperature of $2700$~K \citep{Salz2016}; however, observations from the ground ruled out sub-stellar mass-loss rates higher than $2.5 \times 10^8$~g\,s$^{-1}$ in combination with temperatures lower than $3000$~K \citep{Kasper2020, Spake2022}. For the fiducial simulation of GJ~1214~b, we adopt the escape rate and temperature set by these detection limits.

\begin{figure*}[!ht]
\plotone{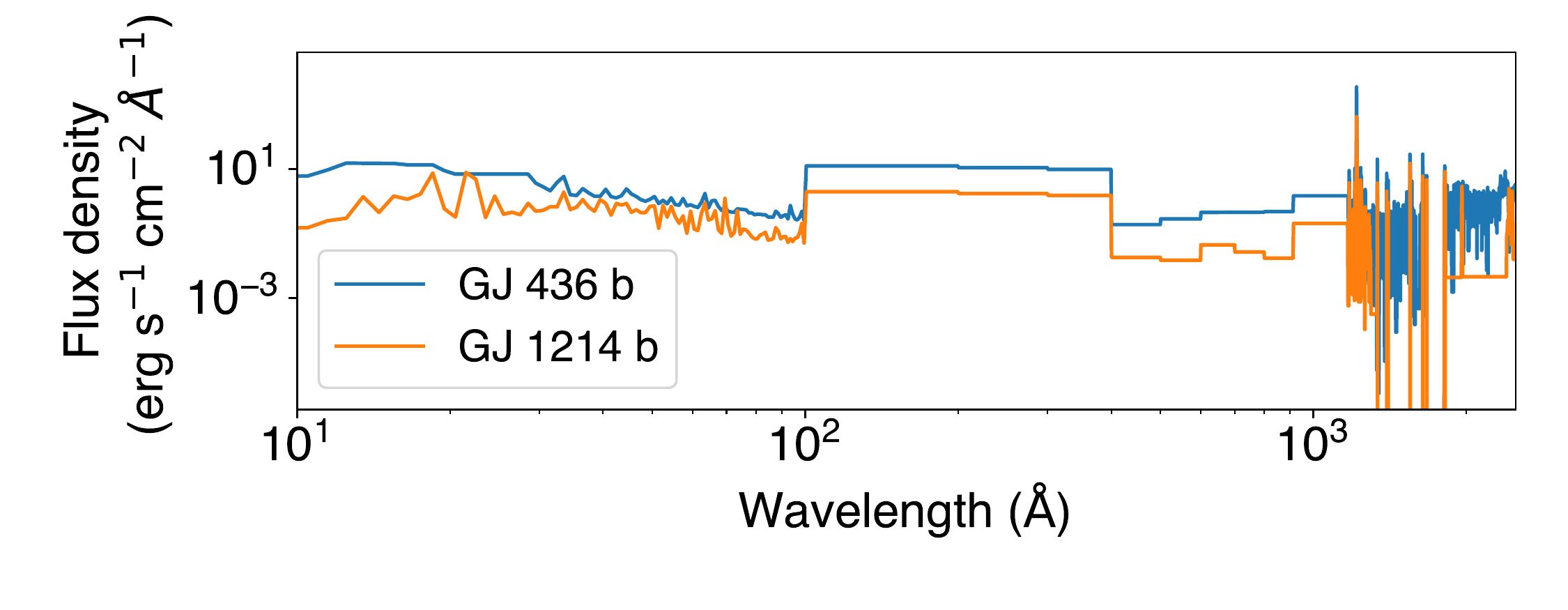}
\caption{High-energy SED incident on GJ~436~b and GJ~1214~b taken from the MUSCLES database and scaled to the semi-major axis of the planets.  \label{fig:gj436_spectrum}}
\end{figure*}

In these exercises, we assume that GJ~436~b and GJ~1214~b have a solar H/He abundance (although we note that their atmospheric compositions have not yet been conclusively constrained by observations). We adopt the spectral energy distribution (SED) of GJ~436 and GJ~1214 measured in the MUSCLES program\footnote{\footnotesize{Publicly available in \url{https://archive.stsci.edu/prepds/muscles/}.}} \citep[][see Figure \ref{fig:gj436_spectrum}]{France2016}. Finally, we assume that the baseline transit depth around the metastable He triplet is constant with a value determined by the average of the transit depth, $(R_{\rm p}/R_{\rm s})^2$, calculated from one-dimensional radiative-convection thermochemical equilibrium forward models from the ATMO model grid \citep{Goyal2019}, scaled to the system parameters for these planets. These forward models assume solar metallicities and C/O and moderate cloud coverage ($\alpha_{cloud}$=0.06). The value of this baseline transit depth does not have a significant impact on the outcomes of our exercises, since the He signal arises from above the planet's thermosphere and is simply added to the transmission spectrum of the lower atmospheric layers.

The final product of the {\tt p-winds} simulation is a theoretical transmission spectrum at infinite spectral resolution. In order to estimate the observable signal, we first need to convolve this theoretical spectrum to the line spread function (LSF) of the instrument, bin the spectrum to its native wavelength grid, and estimate its uncertainties. For the NIRISS/SOSS LSF, we used the same value used to analyze the He signal in HAT-P-18~b presented in \citet{Fu2022}, and which gave good qualitative agreement with the observed signal. For the NIRSpec G140M and G140H modes, we estimated the LSF using an observation from commissioning program PID 1128 (PI: Luetzgendorf; \dataset[DOI: 10.17909/2c5e-dm80]{https://doi.org/10.17909/2c5e-dm80}). In particular, we used the FWHM in the cross-dispersion direction of the observations with G140M and G140H in that program as an estimate of the LSF. To perform that measurement, we followed the same procedures described in \cite{Espinoza2023} to estimate the FWHM as a function of wavelength of NIRSpec/G395H. Briefly, the methodology works with the \texttt{*rateint.fits} JWST pipeline products, which are median-combined, and with which the FWHM is estimated at each column of the spectra under study. Because the spectra are highly tilted and undersampled, the FWHM varies significantly as a function of wavelength. The underlying, physical FWHM of the instrument/mode is estimated as the lower envelope of this FWHM as a function of wavelength curve.



We simulated a single \emph{JWST} observation by binning the signal observable with G140H to a grid of wavelengths from the aforementioned commissioning observation and injecting noise based on a PandExo \citep[version 2.0;][]{Batalha2017} simulation for GJ~436~b. We estimate the significance of the detection by fitting a family of Gaussian profiles using the Markov chain Monte Carlo sampler {\tt emcee} \citep{emcee2013}; the signal-to-noise ratio of the detection is defined as the inferred amplitude of the profile divided by its uncertainty. 

\section{Results}\label{sect:results}

\begin{figure*}[!ht]
\centering
    \gridline{\fig{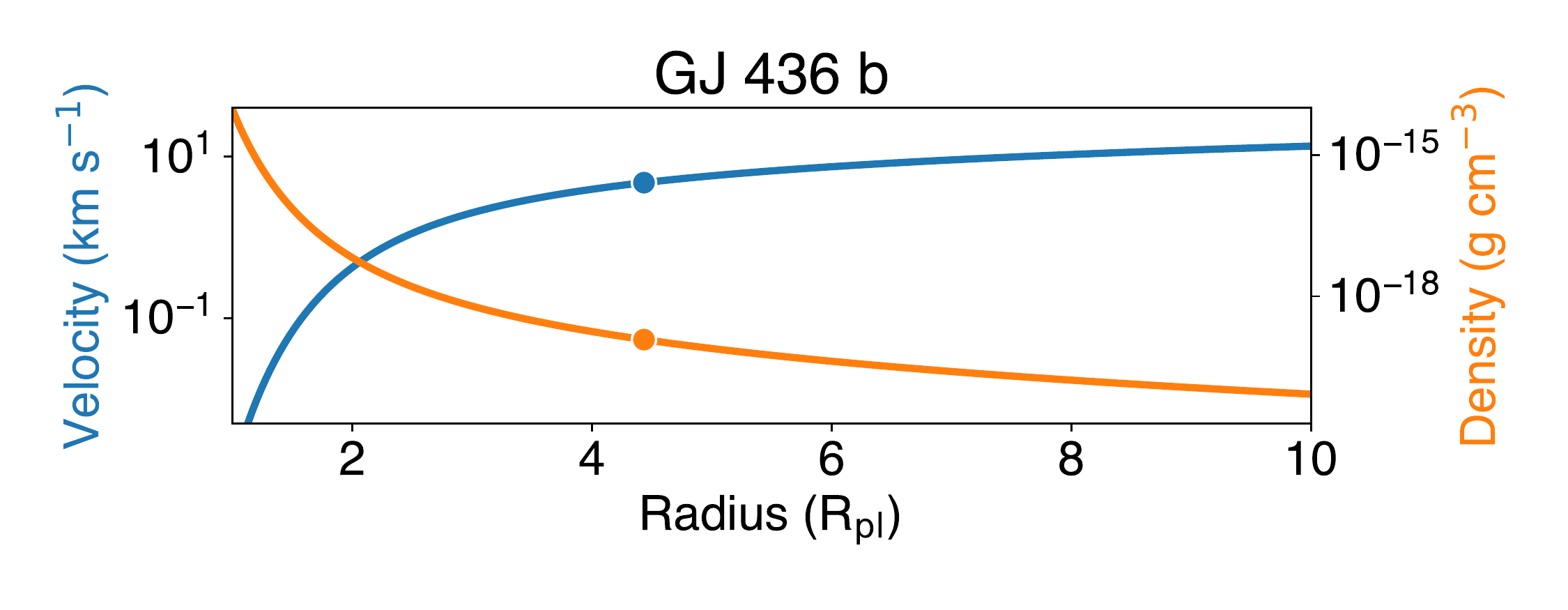}{0.49\textwidth}{(a)}
              \fig{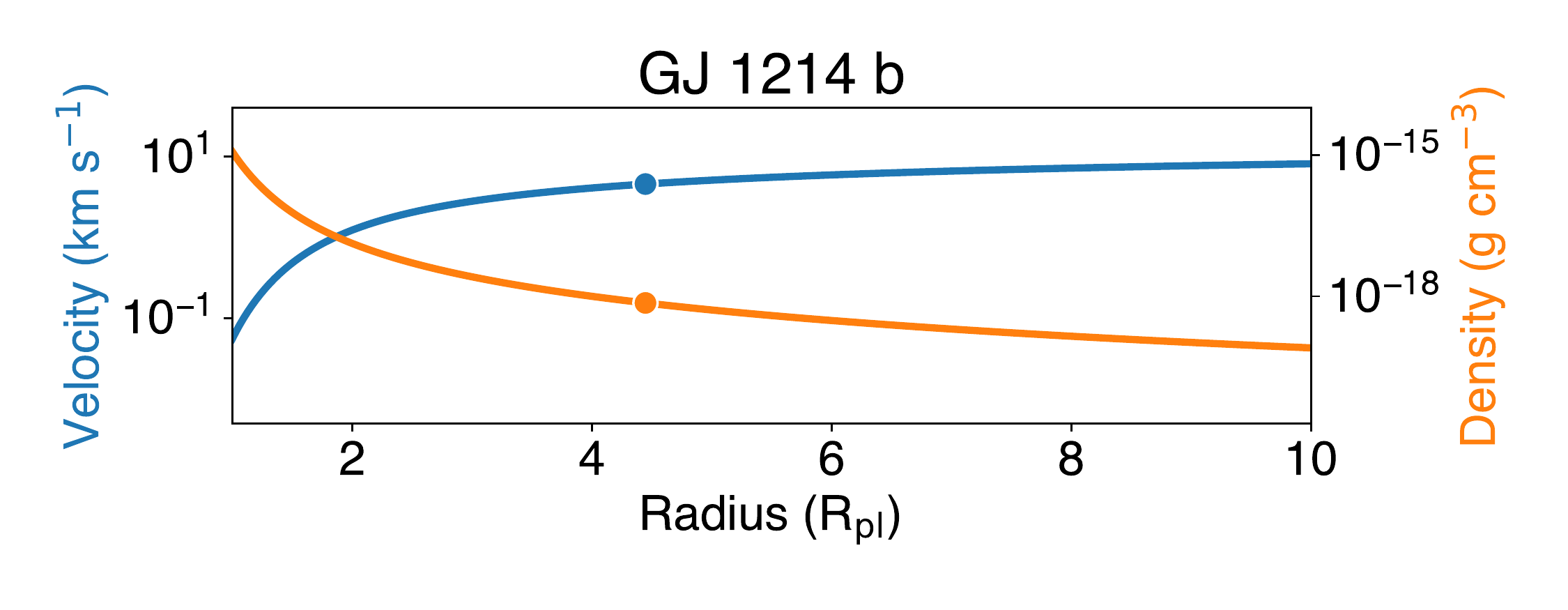}{0.49\textwidth}{(b)}}
    \gridline{\fig{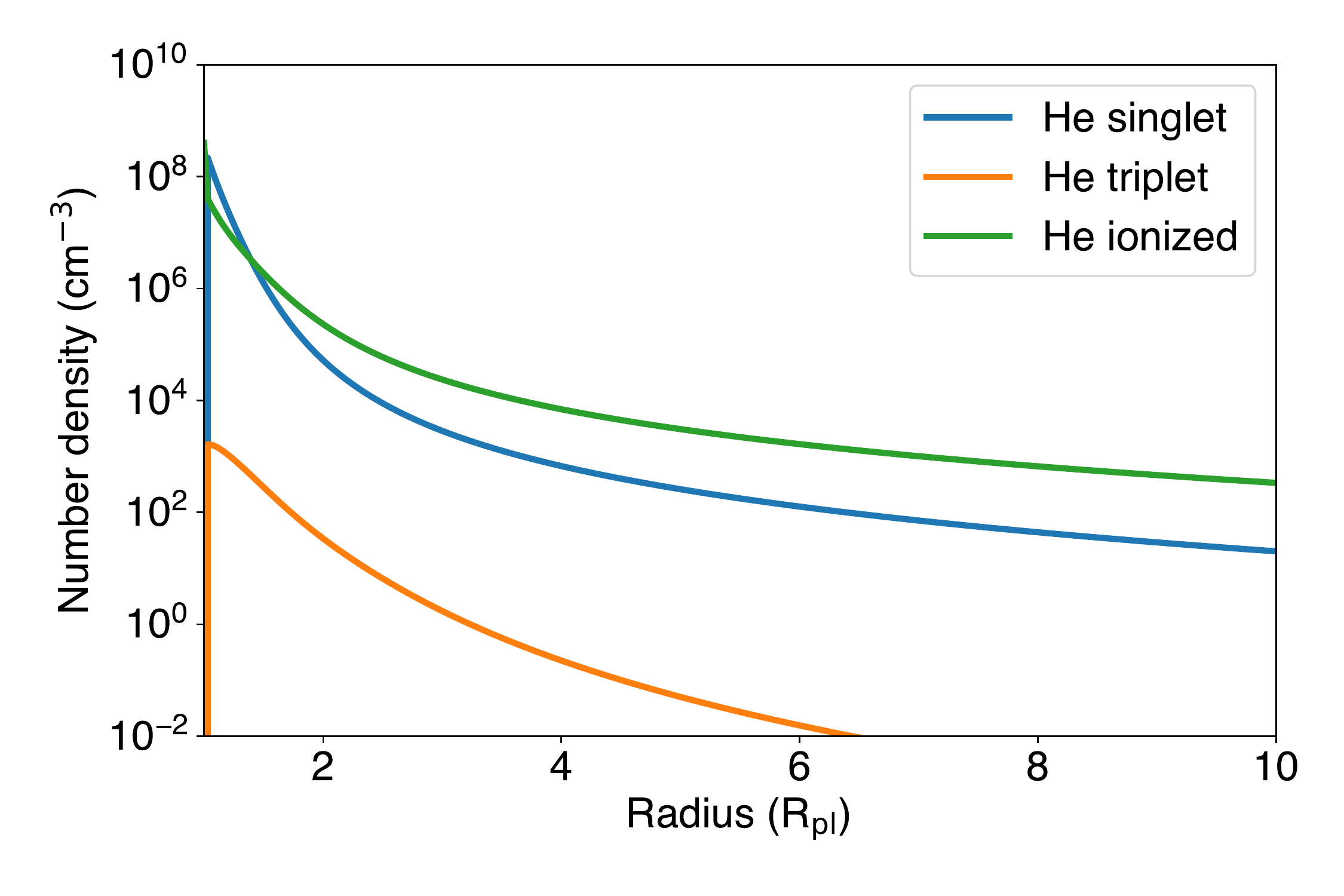}{0.49\textwidth}{(c)}
              \fig{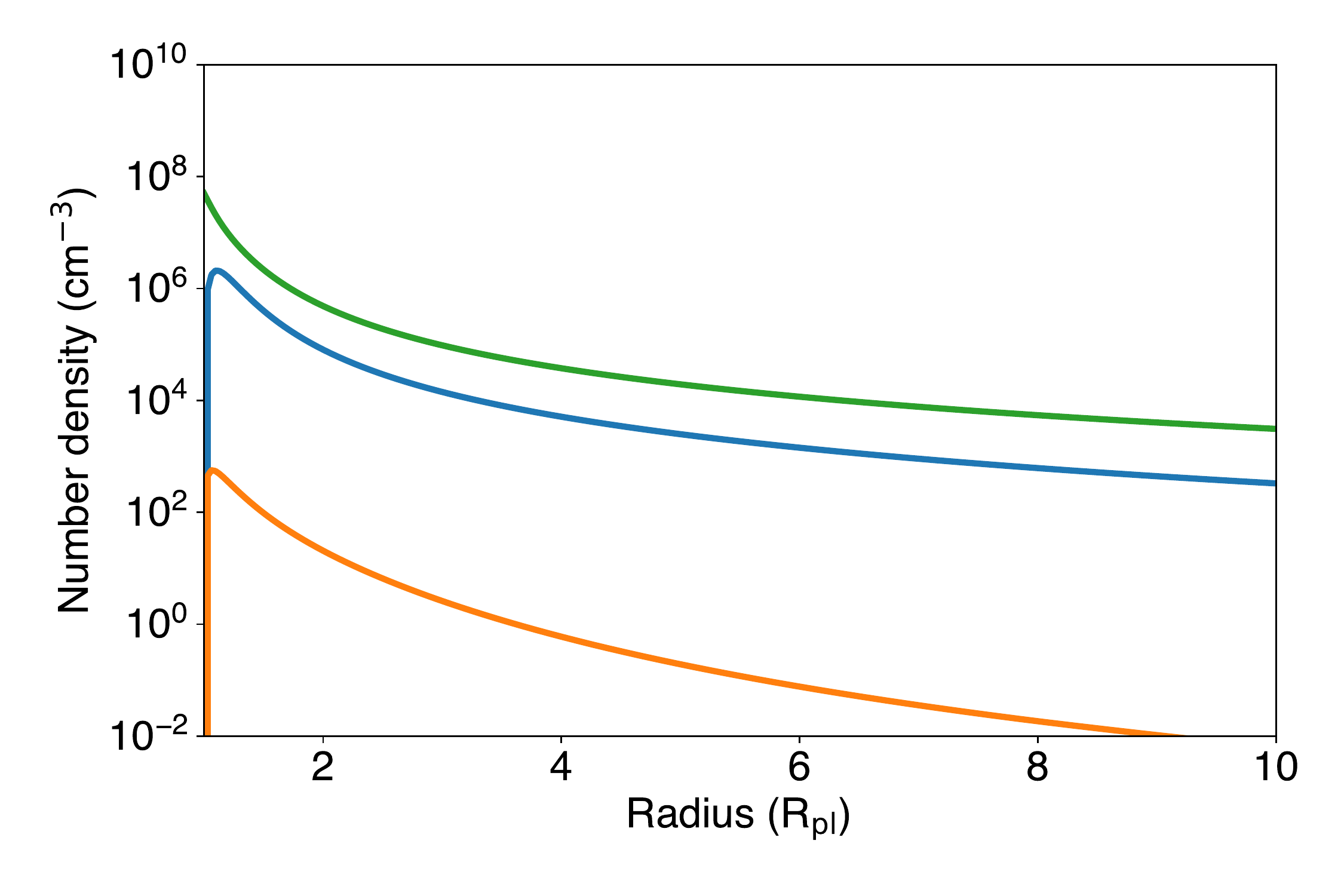}{0.49\textwidth}{(d)}}
\caption{{\it Upper panels:} Structure of the upper atmosphere of GJ~436~b (left) and GJ~1214~b (right). The circles represent the sonic point of the outflow. {\it Lower panels:} Distribution of ionized, singlet and triplet He nuclei in the upper atmosphere of GJ~436~b (left) and GJ~1214~b (right).  \label{fig:model}}
\end{figure*}

\begin{figure*}[!ht]
\centering
\plotone{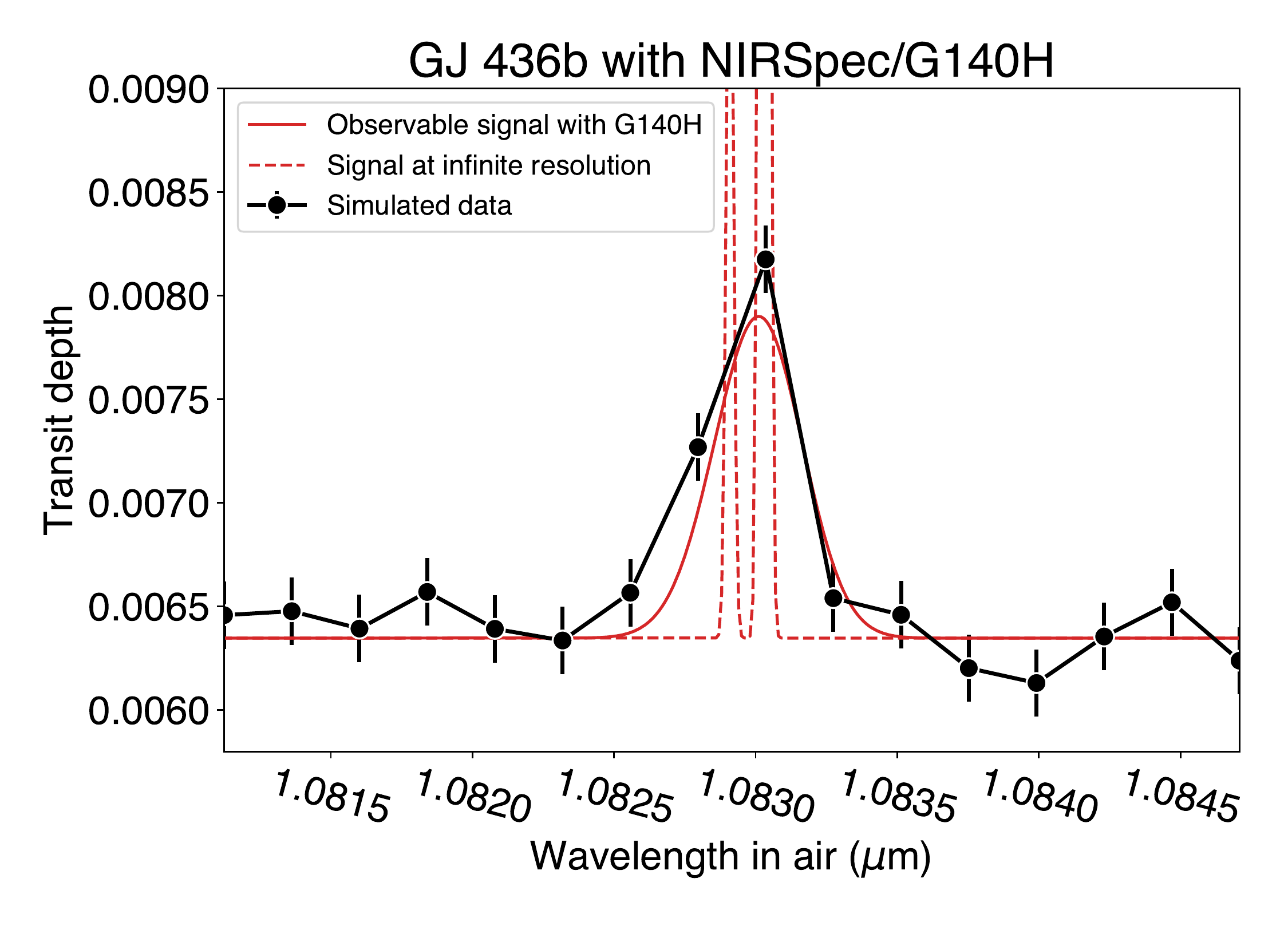}
\caption{Theoretical (red) and observable (black) metastable He transmission spectra of GJ~436~b assuming a sub-stellar escape rate of $10^{8}$~g\,s$^{-1}$ and an outflow temperature of $2700$~K.  \label{fig:tspec_inf_res}}
\end{figure*}

The resulting outflow structures and He distributions for GJ~436~b and GJ~1214~b are shown in Figure \ref{fig:model}, and the predicted transmission spectra at infinite resolution are represented as dashed red curves in Figure \ref{fig:tspec_inf_res}. The metastable He in-transit absorption is narrower than the instrumental LSF, so the signal is spread over 3-4 pixels and assumes the shape of the LSF (see continuous red curves in Figure \ref{fig:tspec_inf_res}).

For GJ~436~b, our simulations indicate that, if the planet has a sub-stellar escape rate of $10^{8}$~g\,s$^{-1}$ and an outflow temperature of $2700$~K, its metastable He signature should be observable with NIRSpec/G140H at $9\sigma$ confidence. For a higher outflow temperature of $3000$~K, the signature will only be marginally detectable in one transit at $\sim 3\sigma$ confidence (see Appendix \ref{app:other}). In general, lower mass loss rates and higher outflow temperatures tend to decrease the detectability of the He signal. Our calculations further show that the in-transit signature is spread over 3-4 pixels. This line broadening is dominated by the instrumental LSF, since the intrinsic broadening of the He signal is narrower than one NIRSpec/G140H pixel (see Figure \ref{fig:tspec_inf_res}).

 We used this simulated transmission spectrum (black symbols in Figure \ref{fig:tspec_inf_res}) to retrieve the escape rate and outflow temperature of GJ~436~b by fitting a family of {\tt p-winds} models using {\tt emcee}. For this retrieval, we set flat priors for both the sub-stellar mass loss rate $\log{\dot{m}}$ and the outflow temperature $\log{T}$. The temperature lower bound is set to $1700$~K, which is the limit where cooler models cannot be calculated due to numerical limitations; this is also just below the $2000$~K lower limit required to thermally dissociate H$_2$ at the base of the wind \citep{MClay2009,Salz2016}. The lower limit for the mass-loss rate was set arbitrarily to $10^7$~g\,s$^{-1}$. The upper limits for temperature and mass-loss rate were set to $7000$~K and $10^{11}$~g\,s$^{-1}$, respectively, by assuming that the outflow is powered by photoionization \citep[see the formulation in][]{Vissapragada2022a}.

\begin{figure}[!ht]
\centering
\plotone{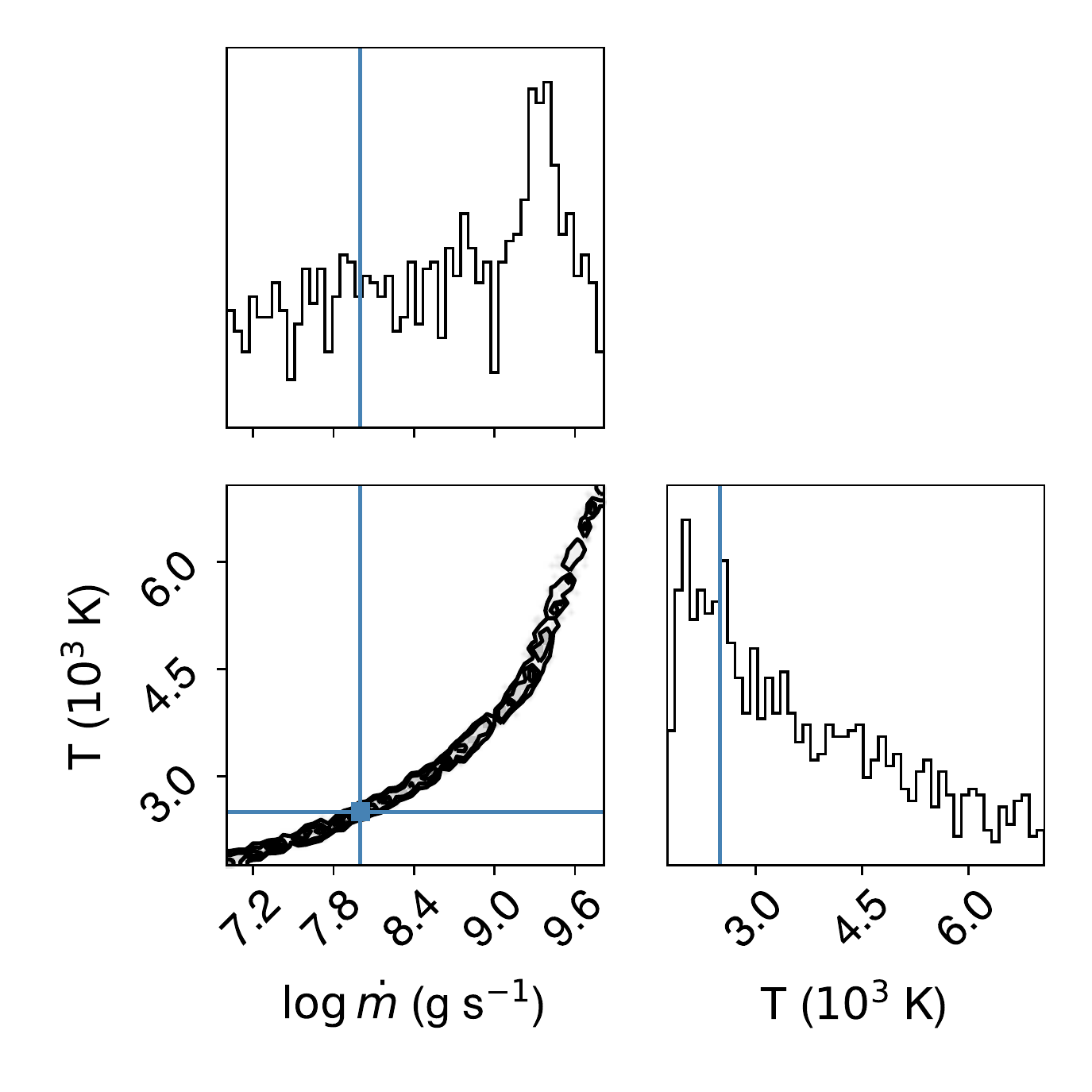}
\caption{Posterior distributions of sub-stellar mass-loss rate and outflow temperature of GJ~436~b in a mock retrieval with {\tt p-winds}. The injected truth is shown as the blue cross-hair.  \label{fig:retrieval}}
\end{figure}

Our mock retrieval for GJ~436~b shows that, even at the highest spectral resolution (R$\sim$2700) of \emph{JWST} with NIRSpec/G140H, our mass loss rate estimates are degenerate with the outflow temperature (see Figure \ref{fig:retrieval}). This degeneracy occurs when the planetary absorption is not spectrally resolved, and it has been observed in studies using narrow-band photometry \citep[e.g.,][]{Vissapragada2022a} and \emph{HST}/WFC3 \citep[e.g.,][]{Mansfield2018}. The reason behind this effect is that the temperature of the outflowing material, which is a free parameter in Parker-wind models, is constrained by measuring the broadening of the planetary absorption. When the absorption is not spectrally resolved, its broadening is dominated by the instrumental LSF, and the temperature remains unconstrained. Since the transit depth is affected by both the outflow temperature and the mass-loss rate of the underlying Parker-wind model, there are many combinations of these two parameters that can fit the observed low-resolution transmission spectrum, hence the degeneracy.

There are different interpretation frameworks that could break this degeneracy. The \texttt{p-winds} code has a module that calculates the maximum heating efficiency of a Parker-wind driven by photoionization that sets upper limits to the escape rate and outflow temperature, helping break this degeneracy for hot Jovian exoplanets \citep[see][]{Vissapragada2022a, Vissapragada2022b}. However, in our simulations we verified that this technique is not very constraining for sub-Jovian worlds at mild levels of irradiation like GJ~436~b and GJ~1214~b. Another technique that aims to limit the parameter space of $\dot{m}$ versus $T$ in unresolved metastable He spectroscopy involves taking into account the effects of radiative heating/cooling, expansion cooling and heat advection \citep[see][]{Linssen2022}. Ideally, one- or three-dimensional hydrodynamics models \citep[such as the ones described in][]{Salz2016, Shaikhislamov2021, MacLeod2022, Kubyshkina2022} break this degeneracy by calculating the outflow temperature profile and mass-loss rates self-consistently, but they are more computationally expensive and they have other free parameters as well -- such as atmospheric abundances, stellar wind strength, and planetary magnetic field strength.

\begin{figure}[!ht]
\centering
\gridline{\fig{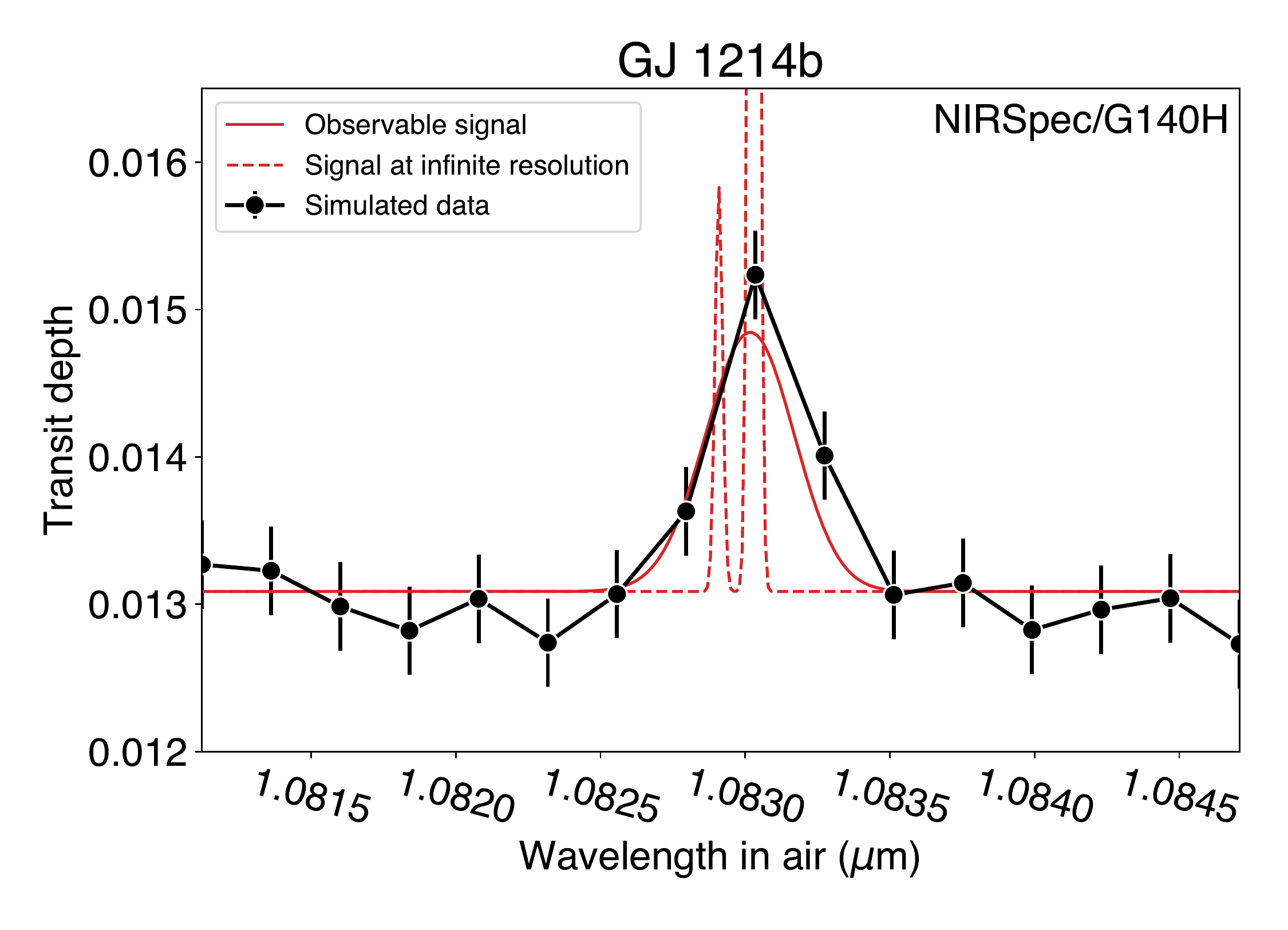}{0.5\textwidth}{}}
\gridline{\fig{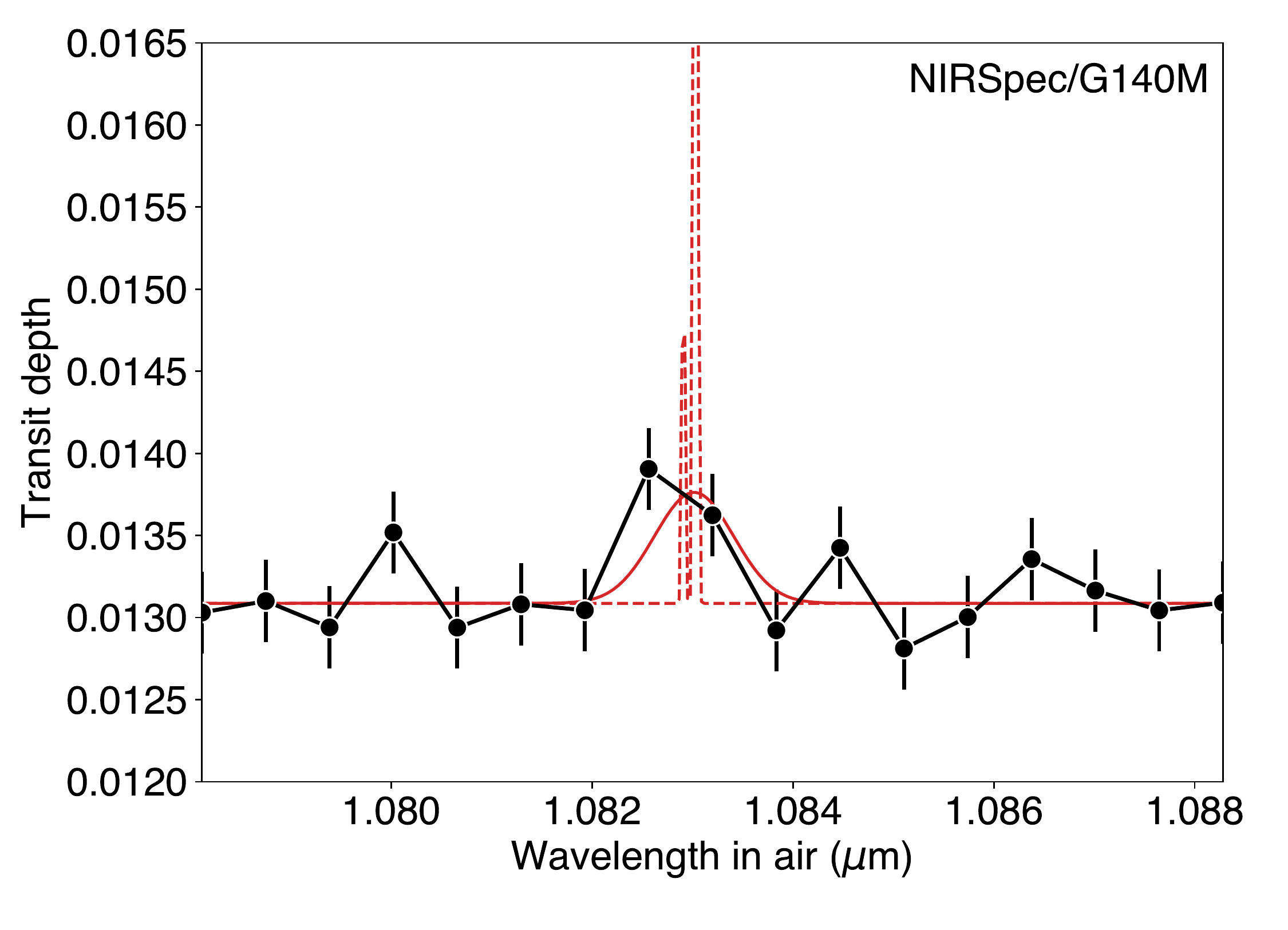}{0.5\textwidth}{}}
\gridline{\fig{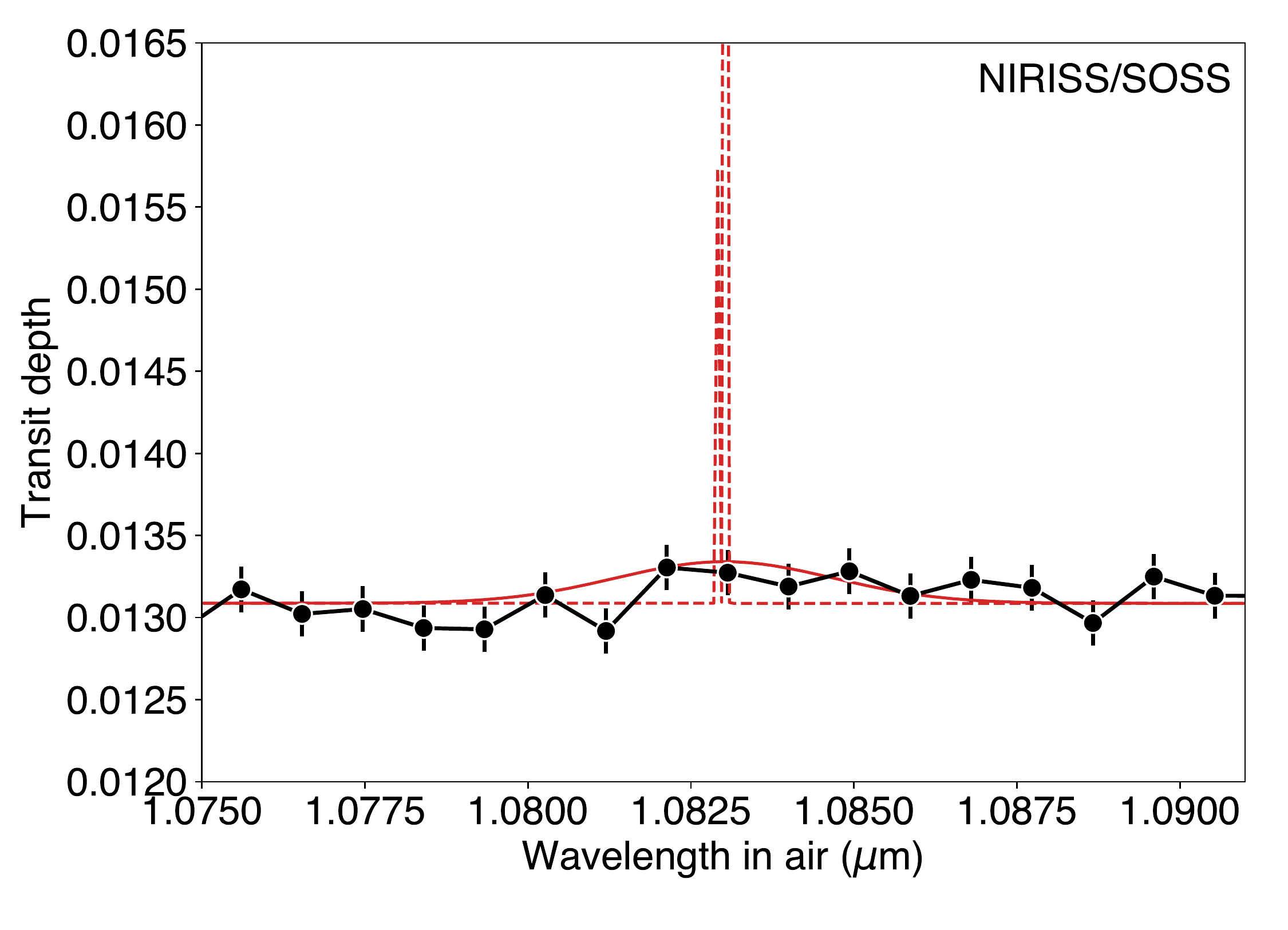}{0.5\textwidth}{}}

\caption{Metastable He transmission spectra of GJ~1214~b simulated for NIRSpec/G140H, G140M and NIRISS/SOSS.  \label{fig:gj1214b_he_spec}}
\end{figure}

Our simulations of the metastable He transmission spectrum of GJ~1214~b show that, for a sub-stellar mass loss rate of $2.5 \times 10^{8}$~g\,s$^{-1}$ and outflow temperature of $T=3000$~K (based on the detection limits of \citealt{Kasper2020}), the signal is detectable with NIRSpec/G140H at $5\sigma$ confidence. The resulting transmission spectra we predict are shown in Figure \ref{fig:gj1214b_he_spec}. These simulated data demonstrate that NIRSpec/G140H yields the best-quality metastable He signal compared to NIRSpec/G140M and NIRISS/SOSS, as it would not be detectable in these other two modes (significance of $\sim 2\sigma$ for first and $\sim 1\sigma$ for the second). The trade-off between more precise fluxes and the narrow He signal spreading over a wider wavelength range is not favorable to detect atmospheric escape. This is not the case, however, for hot Jupiters with strong outflows, such as HAT-P-18~b, whose He signal was detected with narrow-band photometry \citep{Paragas2021} and NIRISS/SOSS \citep{Fu2022}. In fact, during our calculations for GJ~1214~b using the mass-loss rate theoretically predicted by \citet{Salz2016}, the signal would be readily detectable at high confidence with NIRISS/SOSS (see Appendix \ref{app:other}).

\section{Conclusions and recommendations}\label{sect:conclusions}

We simulated the transmission spectra of the warm Neptunes GJ~436~b and GJ~1214~b for a single transit with \emph{JWST} to evaluate the detectability of escaping He and assess how much information we can readily extract from such observations. Our isothermal Parker-wind model of GJ~436~b shows that, for an escape rate of $10^{8}$~g\,s$^{-1}$, an outflow temperature of $2700$~K, and solar atmospheric abundances, the signal is confidently detected with G140H. Such a shallow signal cannot be observed from the ground, where we have access only to higher mass-loss rates ($\dot{m} \gtrsim 10^{10}$~g\,s$^{-1}$). However, the signal size is sensitive to the underlying escape rate and temperature: cooler and stronger outflows are more easily detected, while hotter and weaker outflows are less detectable. 

Our exercises demonstrate that observing atmospheric escape in sub-Jovian worlds using the metastable He line will require \emph{JWST} unless they have strong outflows \citep[see the cases of HAT-P-11~b, GJ~3470~b and some young mini-Neptunes;][]{Allart2018, Palle2020, Zhang2023}. Weaker outflows are expected in planets with lower-than-solar He abundances, metal-rich atmospheres \citep{Ito2021, Nakayama2022}, and those whose upper atmospheres are subject to stellar wind confinement \citep{Vidotto2020, MacLeod2022}.

We generated a simulated transmission spectrum of GJ~436~b with an injected metastable He signal using PandExo. Based on this spectrum, we ran a mock retrieval to estimate the constraints on outflow temperature and mass-loss rate from \emph{JWST} observations. We found that, since the planetary absorption is not spectrally resolved, there is a degeneracy between the retrieved escape rate and temperature when using an isothermal Parker-wind model. 

If we aim to obtain precise constraints on the mass loss rates of sub-Jovian exoplanets with \emph{JWST}, it is crucial to use more sophisticated modeling than the 1D isothermal Parker-wind approximation in order to break the degeneracy between outflow temperature and escape rate. Although \citet{Vissapragada2022a} provides a framework to set upper limits by assuming that the outflow is solely powered by photoionization, these upper limits are not as informative for sub-Jovians as they are for hot gas giants. Self-consistent hydrodynamic simulations do not have this degeneracy, since the outflow temperatures and escape rates are set by the underlying physics of the model \citep[e.g.][]{Salz2016, Shaikhislamov2021, Kubyshkina2022}. The approach described by \citet{Linssen2022} may also be helpful, as they rule out part of the temperature and mass loss parameter space by calculating the temperature structure of the outflow based on a self-consistent photoionization model.

Since GJ~436 can be observed only with NIRSpec/G140H due to saturation, we used a similar mock observation of GJ~1214~b to assess whether the NIRSpec/G140M and NIRISS/SOSS modes are viable options to observe metastable He. These other modes are sometimes desirable because they have a wider wavelength coverage, and yield more information about the atmosphere of the planet than the narrower range accessible with NIRSpec/G140H. However, we found that the other modes are less optimal than G140H to detect escaping He, owing mainly to the signal being diluted over several pixels at lower spectral resolution. \emph{JWST} users wishing to detect atmospheric escape in sub-Jovian worlds should thus use NIRSpec/G140H.

\begin{acknowledgments}
We thank Brett Morris for the coding advice, Nicole Arulanantham for the exchanges about radiative transfer, and the anonymous referee for the helpful review. The {\tt p-winds} code has contributions from Dion Linssen, Lars Klijn, Yassin Jaziri and Michael Gully-Santiago in the form of finding bugs and improving documentation. This project used archival \emph{JWST} data openly available in the Mikulski Archive for Space Telescopes (MAST), which is maintained by the Space Telescope Science Institute (STScI). STScI is operated by the Association of Universities for Research in Astronomy, Inc. under NASA contract NAS 5-26555. This research made use of the NASA Exoplanet Archive, which is operated by the California Institute of Technology, under contract with the National Aeronautics and Space Administration under the Exoplanet Exploration Program.
\end{acknowledgments}

%

\vspace{5mm}
\facilities{JWST(NIRSPec, NIRISS)}


\software{NumPy \citep{harris2020array}, SciPy \citep{2020SciPy-NMeth}, Astropy \citep{astropy:2018}, Jupyter \citep{jupyter}, Matplotlib \citep{Hunter:2007}, {\tt emcee} \citep{emcee2013}, {\tt p-winds} \citep{DSantos2022}, PandExo \citep{Batalha2017}.
      }



\appendix

\section{Other simulations of GJ~436~\lowercase{b} and GJ~1214~\lowercase{b}}\label{app:other}

\begin{figure}[!ht]
\centering
\plotone{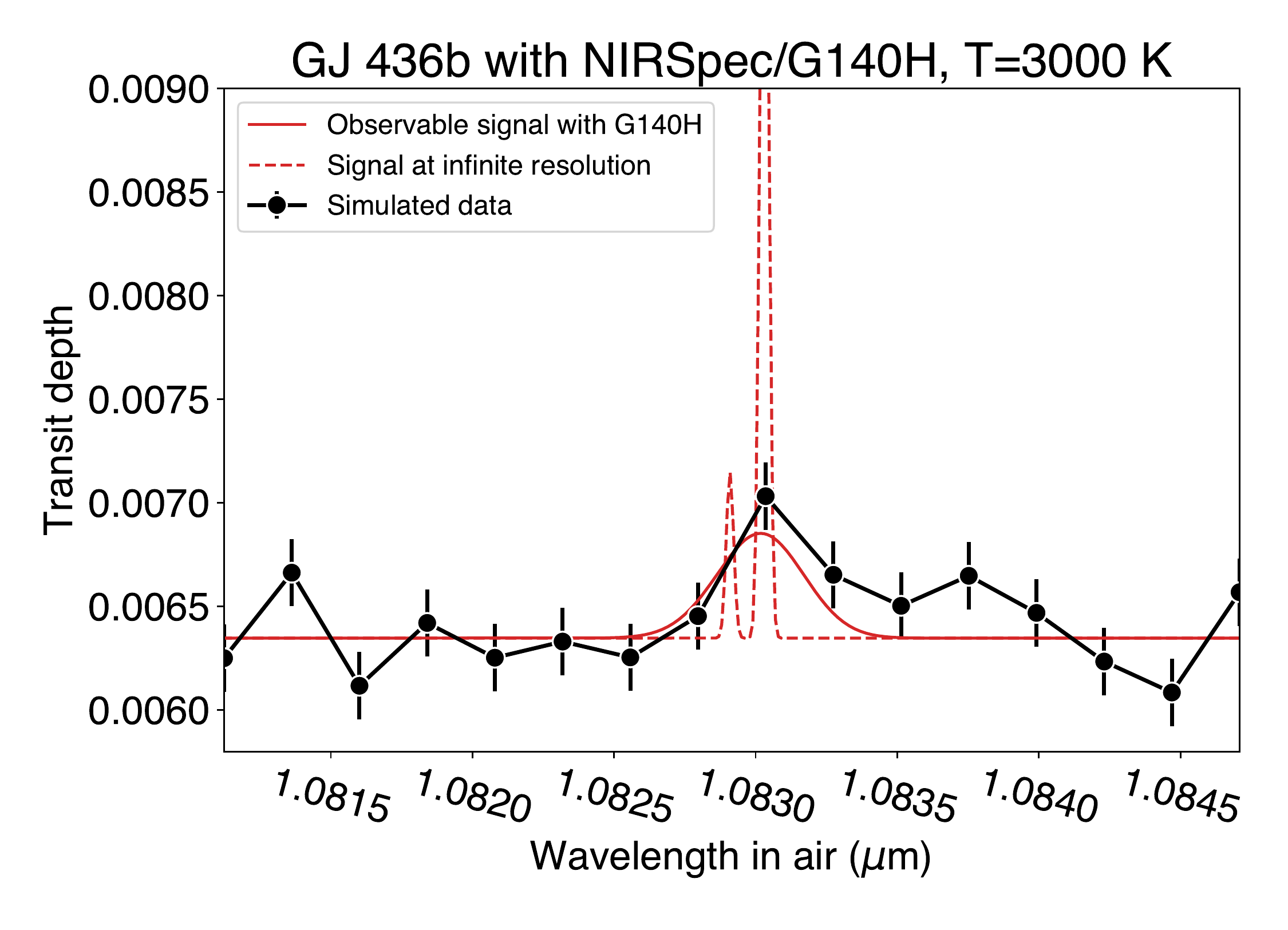}
\caption{Theoretical (red) and observable (black) metastable He transmission spectra of GJ~436~b assuming a sub-stellar escape rate of $10^{8}$~g\,s$^{-1}$ and an outflow temperature of $3000$~K.  \label{fig:tspec_gj436b_3000}}
\end{figure}

We simulated the metastable He signature of GJ~436~b with an escape rate of $\dot{m}=10^8$~g\,s$^{-1}$ and an outflow temperature of $T = 3000$~K to illustrate the effect of higher temperatures being less detectable. The resulting transmission spectrum is shown in Figure \ref{fig:tspec_gj436b_3000}. The signal is only marginally detectable with NIRSpec/G140H with a $\sim 3\sigma$ significance in one transit. However, more transits can be co-added to improve the signal-to-noise ratio of the transmission spectrum.

\begin{figure}[!ht]
\centering
\plotone{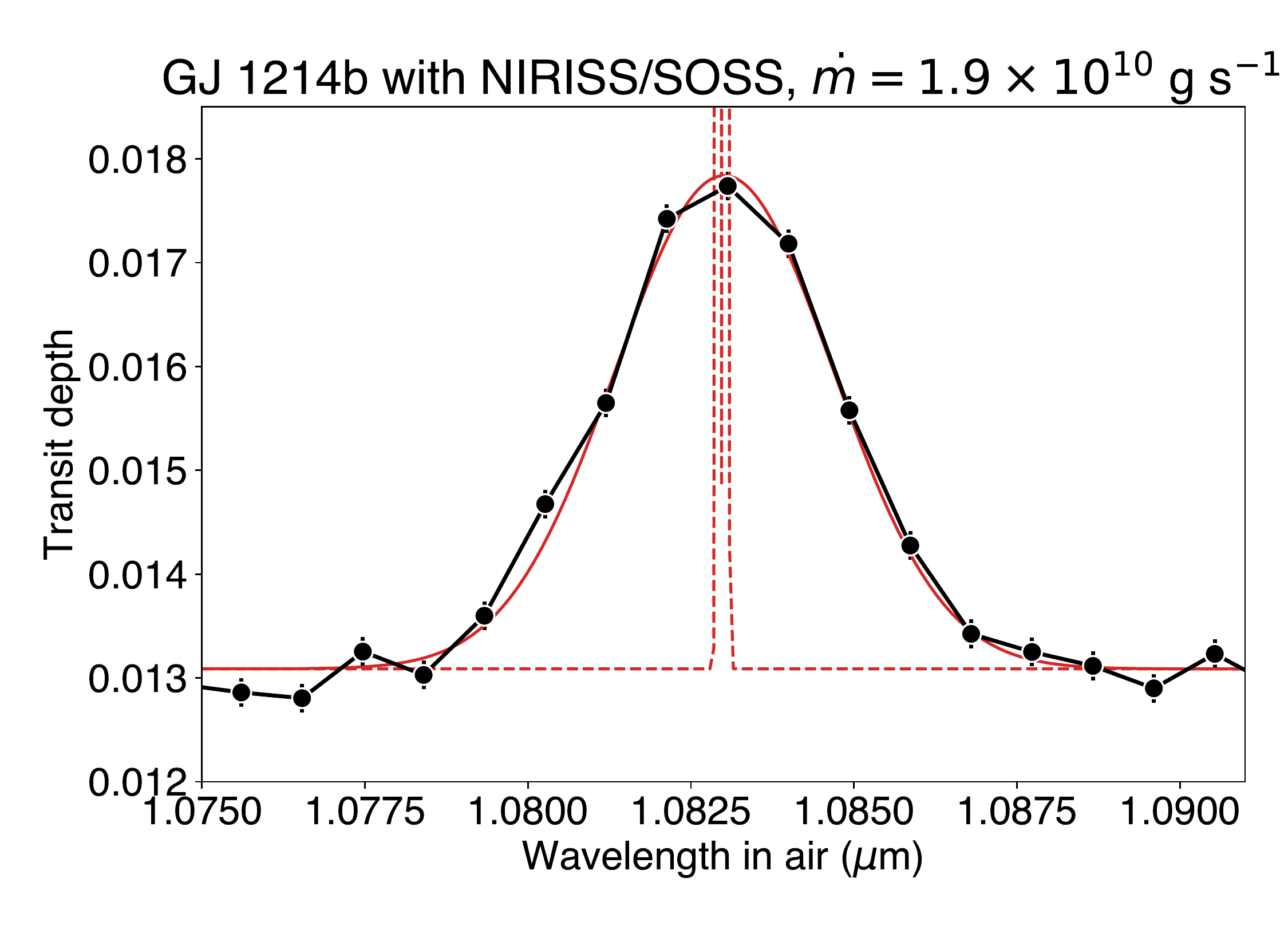}
\caption{Theoretical (red) and observable (black) metastable He transmission spectra of GJ~1214~b assuming a sub-stellar escape rate of $1.9 \times 10^{10}$~g\,s$^{-1}$ and an outflow temperature of $2700$~K based on the self-consistent models of \citet{Salz2016}.  \label{fig:tspec_gj1214b_salz}}
\end{figure}

We also simulated the metastable He signature of GJ~1214~b based on the self-consistent models of \citet{Salz2016}, which predict an escape rate of $\dot{m}=1.9 \times 10^{10}$~g\,s$^{-1}$ and an outflow temperature of $T = 2700$~K. The resulting transmission spectrum is shown in Figure \ref{fig:tspec_gj1214b_salz}. Due to a significantly higher mass-loss rate, this signal would be readily observable with NIRISS/SOSS and even from ground-based observations. The non-detection obtained by \citet{Kasper2020} with Keck~II/NIRSPEC put an upper limit in the mass-loss rate of $\dot{m} < 2.5 \times 10^{8}$~g\,s$^{-1}$.

\bibliography{references}{}

\begin{thebibliography}{}
\expandafter\ifx\csname natexlab\endcsname\relax\def\natexlab#1{#1}\fi
\providecommand{\url}[1]{\href{#1}{#1}}
\providecommand{\dodoi}[1]{doi:~\href{http://doi.org/#1}{\nolinkurl{#1}}}
\providecommand{\doeprint}[1]{\href{http://ascl.net/#1}{\nolinkurl{http://ascl.net/#1}}}
\providecommand{\doarXiv}[1]{\href{https://arxiv.org/abs/#1}{\nolinkurl{https://arxiv.org/abs/#1}}}

\bibitem[{{Allart} {et~al.}(2018){Allart}, {Bourrier}, {Lovis}, {Ehrenreich},
  {Spake}, {Wyttenbach}, {Pino}, {Pepe}, {Sing}, \& {Lecavelier des
  Etangs}}]{Allart2018}
{Allart}, R., {Bourrier}, V., {Lovis}, C., {et~al.} 2018, Science, 362, 1384,
  \dodoi{10.1126/science.aat5879}

\bibitem[{{Astropy Collaboration} {et~al.}(2018){Astropy Collaboration},
  {Price-Whelan}, {Sip{\H{o}}cz}, {G{\"u}nther}, {Lim}, {Crawford}, {Conseil},
  {Shupe}, {Craig}, {Dencheva}, {Ginsburg}, {Vand erPlas}, {Bradley},
  {P{\'e}rez-Su{\'a}rez}, {de Val-Borro}, {Aldcroft}, {Cruz}, {Robitaille},
  {Tollerud}, {Ardelean}, {Babej}, {Bach}, {Bachetti}, {Bakanov}, {Bamford},
  {Barentsen}, {Barmby}, {Baumbach}, {Berry}, {Biscani}, {Boquien}, {Bostroem},
  {Bouma}, {Brammer}, {Bray}, {Breytenbach}, {Buddelmeijer}, {Burke},
  {Calderone}, {Cano Rodr{\'\i}guez}, {Cara}, {Cardoso}, {Cheedella}, {Copin},
  {Corrales}, {Crichton}, {D'Avella}, {Deil}, {Depagne}, {Dietrich}, {Donath},
  {Droettboom}, {Earl}, {Erben}, {Fabbro}, {Ferreira}, {Finethy}, {Fox},
  {Garrison}, {Gibbons}, {Goldstein}, {Gommers}, {Greco}, {Greenfield},
  {Groener}, {Grollier}, {Hagen}, {Hirst}, {Homeier}, {Horton}, {Hosseinzadeh},
  {Hu}, {Hunkeler}, {Ivezi{\'c}}, {Jain}, {Jenness}, {Kanarek}, {Kendrew},
  {Kern}, {Kerzendorf}, {Khvalko}, {King}, {Kirkby}, {Kulkarni}, {Kumar},
  {Lee}, {Lenz}, {Littlefair}, {Ma}, {Macleod}, {Mastropietro}, {McCully},
  {Montagnac}, {Morris}, {Mueller}, {Mumford}, {Muna}, {Murphy}, {Nelson},
  {Nguyen}, {Ninan}, {N{\"o}the}, {Ogaz}, {Oh}, {Parejko}, {Parley}, {Pascual},
  {Patil}, {Patil}, {Plunkett}, {Prochaska}, {Rastogi}, {Reddy Janga},
  {Sabater}, {Sakurikar}, {Seifert}, {Sherbert}, {Sherwood-Taylor}, {Shih},
  {Sick}, {Silbiger}, {Singanamalla}, {Singer}, {Sladen}, {Sooley},
  {Sornarajah}, {Streicher}, {Teuben}, {Thomas}, {Tremblay}, {Turner},
  {Terr{\'o}n}, {van Kerkwijk}, {de la Vega}, {Watkins}, {Weaver}, {Whitmore},
  {Woillez}, {Zabalza}, \& {Astropy Contributors}}]{astropy:2018}
{Astropy Collaboration}, {Price-Whelan}, A.~M., {Sip{\H{o}}cz}, B.~M., {et~al.}
  2018, \aj, 156, 123, \dodoi{10.3847/1538-3881/aabc4f}

\bibitem[{{Ballester} \& {Ben-Jaffel}(2015)}]{Ballester2015}
{Ballester}, G.~E., \& {Ben-Jaffel}, L. 2015, \apj, 804, 116,
  \dodoi{10.1088/0004-637X/804/2/116}

\bibitem[{{Batalha} {et~al.}(2017){Batalha}, {Mandell}, {Pontoppidan},
  {Stevenson}, {Lewis}, {Kalirai}, {Earl}, {Greene}, {Albert}, \&
  {Nielsen}}]{Batalha2017}
{Batalha}, N.~E., {Mandell}, A., {Pontoppidan}, K., {et~al.} 2017, \pasp, 129,
  064501, \dodoi{10.1088/1538-3873/aa65b0}

\bibitem[{{Beaug{\'e}} \& {Nesvorn{\'y}}(2013)}]{Beauge2013}
{Beaug{\'e}}, C., \& {Nesvorn{\'y}}, D. 2013, \apj, 763, 12,
  \dodoi{10.1088/0004-637X/763/1/12}

\bibitem[{{Bourrier} {et~al.}(2016){Bourrier}, {Lecavelier des Etangs},
  {Ehrenreich}, {Tanaka}, \& {Vidotto}}]{Bourrier2016}
{Bourrier}, V., {Lecavelier des Etangs}, A., {Ehrenreich}, D., {Tanaka}, Y.~A.,
  \& {Vidotto}, A.~A. 2016, \aap, 591, A121,
  \dodoi{10.1051/0004-6361/201628362}

\bibitem[{{Bourrier} {et~al.}(2018){Bourrier}, {Lecavelier des Etangs},
  {Ehrenreich}, {Sanz-Forcada}, {Allart}, {Ballester}, {Buchhave}, {Cohen},
  {Deming}, {Evans}, {Garc{\'{\i}}a Mu{\~n}oz}, {Henry}, {Kataria}, {Lavvas},
  {Lewis}, {L{\'o}pez-Morales}, {Marley}, {Sing}, \& {Wakeford}}]{Bourrier2018}
{Bourrier}, V., {Lecavelier des Etangs}, A., {Ehrenreich}, D., {et~al.} 2018,
  \aap, 620, A147, \dodoi{10.1051/0004-6361/201833675}

\bibitem[{{Carleo} {et~al.}(2021){Carleo}, {Youngblood}, {Redfield}, {Casasayas
  Barris}, {Ayres}, {Vannier}, {Fossati}, {Palle}, {Livingston}, {Lanza},
  {Niraula}, {Alvarado-G{\'o}mez}, {Chen}, {Gandolfi}, {Guenther}, {Linsky},
  {Nagel}, {Narita}, {Nortmann}, {Shkolnik}, \& {Stangret}}]{Carleo2021}
{Carleo}, I., {Youngblood}, A., {Redfield}, S., {et~al.} 2021, \aj, 161, 136,
  \dodoi{10.3847/1538-3881/abdb2f}

\bibitem[{{Chadney} {et~al.}(2015){Chadney}, {Galand}, {Unruh}, {Koskinen}, \&
  {Sanz-Forcada}}]{Chadney2015}
{Chadney}, J.~M., {Galand}, M., {Unruh}, Y.~C., {Koskinen}, T.~T., \&
  {Sanz-Forcada}, J. 2015, \icarus, 250, 357,
  \dodoi{10.1016/j.icarus.2014.12.012}

\bibitem[{{Chamberlain}(1963)}]{Chamberlain1963}
{Chamberlain}, J.~W. 1963, \planss, 11, 901,
  \dodoi{10.1016/0032-0633(63)90122-3}

\bibitem[{{Dos Santos}(2022)}]{DSantos2023}
{Dos Santos}, L.~A. 2022, arXiv e-prints, arXiv:2211.16243.
\newblock \doarXiv{2211.16243}

\bibitem[{{Dos Santos} {et~al.}(2019){Dos Santos}, {Ehrenreich}, {Bourrier},
  {Lecavelier des Etangs}, {L{\'o}pez-Morales}, {Sing}, {Ballester},
  {Ben-Jaffel}, {Buchhave}, {Garc{\'\i}a Mu{\~n}oz}, {Henry}, {Kataria},
  {Lavie}, {Lavvas}, {Lewis}, {Mikal-Evans}, {Sanz-Forcada}, \&
  {Wakeford}}]{DSantos2019}
{Dos Santos}, L.~A., {Ehrenreich}, D., {Bourrier}, V., {et~al.} 2019, \aap,
  629, A47, \dodoi{10.1051/0004-6361/201935663}

\bibitem[{{Dos Santos} {et~al.}(2022){Dos Santos}, {Vidotto}, {Vissapragada},
  {Alam}, {Allart}, {Bourrier}, {Kirk}, {Seidel}, \&
  {Ehrenreich}}]{DSantos2022}
{Dos Santos}, L.~A., {Vidotto}, A.~A., {Vissapragada}, S., {et~al.} 2022, \aap,
  659, A62, \dodoi{10.1051/0004-6361/202142038}

\bibitem[{{Ehrenreich} {et~al.}(2015){Ehrenreich}, {Bourrier}, {Wheatley},
  {Lecavelier des Etangs}, {H{\'e}brard}, {Udry}, {Bonfils}, {Delfosse},
  {D{\'e}sert}, {Sing}, \& {Vidal-Madjar}}]{Ehrenreich2015}
{Ehrenreich}, D., {Bourrier}, V., {Wheatley}, P.~J., {et~al.} 2015, \nat, 522,
  459, \dodoi{10.1038/nature14501}

\bibitem[{{Erkaev} {et~al.}(2015){Erkaev}, {Lammer}, {Odert}, {Kulikov}, \&
  {Kislyakova}}]{Erkaev2015}
{Erkaev}, N.~V., {Lammer}, H., {Odert}, P., {Kulikov}, Y.~N., \& {Kislyakova},
  K.~G. 2015, \mnras, 448, 1916, \dodoi{10.1093/mnras/stv130}

\bibitem[{{Espinoza} {et~al.}(2022){Espinoza}, {{\'U}beda}, {Birkmann},
  {Ferruit}, {Valenti}, {Sing}, {Rustamkulov}, {Regan}, {Kendrew}, {Sabbi},
  {Schlawin}, {Beatty}, {Albert}, {Greene}, {Nikolov}, {Karakla}, {Keyes},
  {Kumari}, {Alves de Oliveira}, {B{\"o}ker}, {Pe{\~n}a-Guerrero}, {Giardino},
  {Manjavacas}, {Proffitt}, \& {Rawle}}]{Espinoza2023}
{Espinoza}, N., {{\'U}beda}, L., {Birkmann}, S.~M., {et~al.} 2022, arXiv
  e-prints, arXiv:2211.01459, \dodoi{10.48550/arXiv.2211.01459}

\bibitem[{{Foreman-Mackey} {et~al.}(2013){Foreman-Mackey}, {Hogg}, {Lang}, \&
  {Goodman}}]{emcee2013}
{Foreman-Mackey}, D., {Hogg}, D.~W., {Lang}, D., \& {Goodman}, J. 2013, \pasp,
  125, 306, \dodoi{10.1086/670067}

\bibitem[{{Fossati} {et~al.}(2010){Fossati}, {Haswell}, {Froning}, {Hebb},
  {Holmes}, {Kolb}, {Helling}, {Carter}, {Wheatley}, {Collier Cameron},
  {Loeillet}, {Pollacco}, {Street}, {Stempels}, {Simpson}, {Udry}, {Joshi},
  {West}, {Skillen}, \& {Wilson}}]{Fossati2010}
{Fossati}, L., {Haswell}, C.~A., {Froning}, C.~S., {et~al.} 2010, \apjl, 714,
  L222, \dodoi{10.1088/2041-8205/714/2/L222}

\bibitem[{{Fossati} {et~al.}(2022){Fossati}, {Guilluy}, {Shaikhislamov},
  {Carleo}, {Borsa}, {Bonomo}, {Giacobbe}, {Rainer}, {Cecchi-Pestellini},
  {Khodachenko}, {Efimov}, {Rumenskikh}, {Miroshnichenko}, {Berezutsky},
  {Nascimbeni}, {Brogi}, {Lanza}, {Mancini}, {Affer}, {Benatti}, {Biazzo},
  {Bignamini}, {Carosati}, {Claudi}, {Cosentino}, {Covino}, {Desidera},
  {Fiorenzano}, {Harutyunyan}, {Maggio}, {Malavolta}, {Maldonado}, {Micela},
  {Molinari}, {Pagano}, {Pedani}, {Piotto}, {Poretti}, {Scandariato},
  {Sozzetti}, \& {Stoev}}]{Fossati2022}
{Fossati}, L., {Guilluy}, G., {Shaikhislamov}, I.~F., {et~al.} 2022, \aap, 658,
  A136, \dodoi{10.1051/0004-6361/202142336}

\bibitem[{{France} {et~al.}(2016){France}, {Loyd}, {Youngblood}, {Brown},
  {Schneider}, {Hawley}, {Froning}, {Linsky}, {Roberge}, {Buccino},
  {Davenport}, {Fontenla}, {Kaltenegger}, {Kowalski}, {Mauas}, {Miguel},
  {Redfield}, {Rugheimer}, {Tian}, {Vieytes}, {Walkowicz}, \&
  {Weisenburger}}]{France2016}
{France}, K., {Loyd}, R.~O.~P., {Youngblood}, A., {et~al.} 2016, \apj, 820, 89,
  \dodoi{10.3847/0004-637X/820/2/89}

\bibitem[{{Fu} {et~al.}(2022){Fu}, {Espinoza}, {Sing}, {Lothringer}, {Dos
  Santos}, {Rustamkulov}, {Deming}, {Kempton}, {Komacek}, {Knutson}, {Albert},
  {Pontoppidan}, {Volk}, \& {Filippazzo}}]{Fu2022}
{Fu}, G., {Espinoza}, N., {Sing}, D.~K., {et~al.} 2022, \apjl, 940, L35,
  \dodoi{10.3847/2041-8213/ac9977}

\bibitem[{{Fulton} \& {Petigura}(2018)}]{Fulton2018}
{Fulton}, B.~J., \& {Petigura}, E.~A. 2018, \aj, 156, 264,
  \dodoi{10.3847/1538-3881/aae828}

\bibitem[{{Fulton} {et~al.}(2017){Fulton}, {Petigura}, {Howard}, {Isaacson},
  {Marcy}, {Cargile}, {Hebb}, {Weiss}, {Johnson}, {Morton}, {Sinukoff},
  {Crossfield}, \& {Hirsch}}]{Fulton2017}
{Fulton}, B.~J., {Petigura}, E.~A., {Howard}, A.~W., {et~al.} 2017, \aj, 154,
  109, \dodoi{10.3847/1538-3881/aa80eb}

\bibitem[{{Gaidos} {et~al.}(2021){Gaidos}, {Hirano}, {Omiya}, {Kuzuhara},
  {Kotani}, {Tamura}, {Harakawa}, \& {Kudo}}]{Gaidos2021}
{Gaidos}, E., {Hirano}, T., {Omiya}, M., {et~al.} 2021, Research Notes of the
  American Astronomical Society, 5, 238, \dodoi{10.3847/2515-5172/ac31bd}

\bibitem[{{Gaidos} {et~al.}(2020){Gaidos}, {Hirano}, {Mann}, {Owens}, {Berger},
  {France}, {Vanderburg}, {Harakawa}, {Hodapp}, {Ishizuka}, {Jacobson},
  {Konishi}, {Kotani}, {Kudo}, {Kurokawa}, {Kuzuhara}, {Nishikawa}, {Omiya},
  {Serizawa}, {Tamura}, \& {Ueda}}]{Gaidos2020}
{Gaidos}, E., {Hirano}, T., {Mann}, A.~W., {et~al.} 2020, \mnras, 495, 650,
  \dodoi{10.1093/mnras/staa918}

\bibitem[{{Garc{\'\i}a Mu{\~n}oz} {et~al.}(2021){Garc{\'\i}a Mu{\~n}oz},
  {Fossati}, {Youngblood}, {Nettelmann}, {Gandolfi}, {Cabrera}, \&
  {Rauer}}]{Garcia2021}
{Garc{\'\i}a Mu{\~n}oz}, A., {Fossati}, L., {Youngblood}, A., {et~al.} 2021,
  \apjl, 907, L36, \dodoi{10.3847/2041-8213/abd9b8}

\bibitem[{{Ginzburg} {et~al.}(2018){Ginzburg}, {Schlichting}, \&
  {Sari}}]{Ginzburg2018}
{Ginzburg}, S., {Schlichting}, H.~E., \& {Sari}, R. 2018, \mnras, 476, 759,
  \dodoi{10.1093/mnras/sty290}

\bibitem[{{Goyal} {et~al.}(2019){Goyal}, {Wakeford}, {Mayne}, {Lewis},
  {Drummond}, \& {Sing}}]{Goyal2019}
{Goyal}, J.~M., {Wakeford}, H.~R., {Mayne}, N.~J., {et~al.} 2019, \mnras, 482,
  4503, \dodoi{10.1093/mnras/sty3001}

\bibitem[{{Gupta} \& {Schlichting}(2019)}]{Gupta2019}
{Gupta}, A., \& {Schlichting}, H.~E. 2019, \mnras, 487, 24,
  \dodoi{10.1093/mnras/stz1230}

\bibitem[{Harris {et~al.}(2020)Harris, Millman, van~der Walt, Gommers,
  Virtanen, Cournapeau, Wieser, Taylor, Berg, Smith, Kern, Picus, Hoyer, van
  Kerkwijk, Brett, Haldane, del R{\'{i}}o, Wiebe, Peterson,
  G{\'{e}}rard-Marchant, Sheppard, Reddy, Weckesser, Abbasi, Gohlke, \&
  Oliphant}]{harris2020array}
Harris, C.~R., Millman, K.~J., van~der Walt, S.~J., {et~al.} 2020, Nature, 585,
  357, \dodoi{10.1038/s41586-020-2649-2}

\bibitem[{{Hirano} {et~al.}(2020){Hirano}, {Krishnamurthy}, {Gaidos},
  {Flewelling}, {Mann}, {Narita}, {Plavchan}, {Kotani}, {Tamura}, {Harakawa},
  {Hodapp}, {Ishizuka}, {Jacobson}, {Konishi}, {Kudo}, {Kurokawa}, {Kuzuhara},
  {Nishikawa}, {Omiya}, {Serizawa}, {Ueda}, \& {Vievard}}]{Hirano2020}
{Hirano}, T., {Krishnamurthy}, V., {Gaidos}, E., {et~al.} 2020, \apjl, 899,
  L13, \dodoi{10.3847/2041-8213/aba6eb}

\bibitem[{{Hunten} {et~al.}(1987){Hunten}, {Pepin}, \& {Walker}}]{Hunten1987}
{Hunten}, D.~M., {Pepin}, R.~O., \& {Walker}, J.~C.~G. 1987, \icarus, 69, 532,
  \dodoi{10.1016/0019-1035(87)90022-4}

\bibitem[{Hunter(2007)}]{Hunter:2007}
Hunter, J.~D. 2007, Computing in Science \& Engineering, 9, 90,
  \dodoi{10.1109/MCSE.2007.55}

\bibitem[{{Ionov} {et~al.}(2018){Ionov}, {Pavlyuchenkov}, \&
  {Shematovich}}]{Ionov2018}
{Ionov}, D.~E., {Pavlyuchenkov}, Y.~N., \& {Shematovich}, V.~I. 2018, \mnras,
  476, 5639, \dodoi{10.1093/mnras/sty626}

\bibitem[{{Ito} \& {Ikoma}(2021)}]{Ito2021}
{Ito}, Y., \& {Ikoma}, M. 2021, \mnras, 502, 750,
  \dodoi{10.1093/mnras/staa3962}

\bibitem[{{Kasper} {et~al.}(2020){Kasper}, {Bean}, {Oklop{\v{c}}i{\'c}},
  {Malsky}, {Kempton}, {D{\'e}sert}, {Rogers}, \& {Mansfield}}]{Kasper2020}
{Kasper}, D., {Bean}, J.~L., {Oklop{\v{c}}i{\'c}}, A., {et~al.} 2020, \aj, 160,
  258, \dodoi{10.3847/1538-3881/abbee6}

\bibitem[{{Kirk} {et~al.}(2022){Kirk}, {Dos Santos}, {L{\'o}pez-Morales},
  {Alam}, {Oklop{\v{c}}i{\'c}}, {MacLeod}, {Zeng}, \& {Zhou}}]{Kirk2022}
{Kirk}, J., {Dos Santos}, L.~A., {L{\'o}pez-Morales}, M., {et~al.} 2022, \aj,
  164, 24, \dodoi{10.3847/1538-3881/ac722f}

\bibitem[{Kluyver {et~al.}(2016)Kluyver, Ragan-Kelley, P{\'e}rez, Granger,
  Bussonnier, Frederic, Kelley, Hamrick, Grout, Corlay, Ivanov, Avila, Abdalla,
  Willing, \& development team}]{jupyter}
Kluyver, T., Ragan-Kelley, B., P{\'e}rez, F., {et~al.} 2016, in Positioning and
  Power in Academic Publishing: Players, Agents and Agendas, ed. F.~Loizides \&
  B.~Scmidt (Netherlands: IOS Press), 87--90.
\newblock \url{https://eprints.soton.ac.uk/403913/}

\bibitem[{{Kubyshkina} {et~al.}(2022){Kubyshkina}, {Vidotto}, {Villarreal
  D'Angelo}, {Carolan}, {Hazra}, \& {Carleo}}]{Kubyshkina2022}
{Kubyshkina}, D., {Vidotto}, A.~A., {Villarreal D'Angelo}, C., {et~al.} 2022,
  \mnras, 510, 2111, \dodoi{10.1093/mnras/stab3594}

\bibitem[{{Lammer} {et~al.}(2003){Lammer}, {Selsis}, {Ribas}, {Guinan},
  {Bauer}, \& {Weiss}}]{Lammer2003}
{Lammer}, H., {Selsis}, F., {Ribas}, I., {et~al.} 2003, \apjl, 598, L121,
  \dodoi{10.1086/380815}

\bibitem[{{Lamp{\'o}n} {et~al.}(2020){Lamp{\'o}n}, {L{\'o}pez-Puertas}, {Lara},
  {S{\'a}nchez-L{\'o}pez}, {Salz}, {Czesla}, {Sanz-Forcada}, {Molaverdikhani},
  {Alonso-Floriano}, {Nortmann}, {Caballero}, {Bauer}, {Pall{\'e}}, {Montes},
  {Quirrenbach}, {Nagel}, {Ribas}, {Reiners}, \& {Amado}}]{Lampon2020}
{Lamp{\'o}n}, M., {L{\'o}pez-Puertas}, M., {Lara}, L.~M., {et~al.} 2020, \aap,
  636, A13, \dodoi{10.1051/0004-6361/201937175}

\bibitem[{{Lavie} {et~al.}(2017){Lavie}, {Ehrenreich}, {Bourrier}, {Lecavelier
  des Etangs}, {Vidal-Madjar}, {Delfosse}, {Gracia Berna}, {Heng}, {Thomas},
  {Udry}, \& {Wheatley}}]{Lavie2017}
{Lavie}, B., {Ehrenreich}, D., {Bourrier}, V., {et~al.} 2017, \aap, 605, L7,
  \dodoi{10.1051/0004-6361/201731340}

\bibitem[{{Lecavelier des Etangs} {et~al.}(2010){Lecavelier des Etangs},
  {Ehrenreich}, {Vidal-Madjar}, {Ballester}, {D{\'e}sert}, {Ferlet},
  {H{\'e}brard}, {Sing}, {Tchakoumegni}, \& {Udry}}]{Lecavelier2010}
{Lecavelier des Etangs}, A., {Ehrenreich}, D., {Vidal-Madjar}, A., {et~al.}
  2010, \aap, 514, A72, \dodoi{10.1051/0004-6361/200913347}

\bibitem[{{Linsky} {et~al.}(2010){Linsky}, {Yang}, {France}, {Froning},
  {Green}, {Stocke}, \& {Osterman}}]{Linsky2010}
{Linsky}, J.~L., {Yang}, H., {France}, K., {et~al.} 2010, \apj, 717, 1291,
  \dodoi{10.1088/0004-637X/717/2/1291}

\bibitem[{{Linssen} {et~al.}(2022){Linssen}, {Oklop{\v{c}}i{\'c}}, \&
  {MacLeod}}]{Linssen2022}
{Linssen}, D.~C., {Oklop{\v{c}}i{\'c}}, A., \& {MacLeod}, M. 2022, \aap, 667,
  A54, \dodoi{10.1051/0004-6361/202243830}

\bibitem[{{MacLeod} \& {Oklop{\v{c}}i{\'c}}(2022)}]{MacLeod2022}
{MacLeod}, M., \& {Oklop{\v{c}}i{\'c}}, A. 2022, \apj, 926, 226,
  \dodoi{10.3847/1538-4357/ac46ce}

\bibitem[{{Mansfield} {et~al.}(2018){Mansfield}, {Bean}, {Oklop{\v{c}}i{\'c}},
  {Kreidberg}, {D{\'e}sert}, {Kempton}, {Line}, {Fortney}, {Henry}, {Mallonn},
  {Stevenson}, {Dragomir}, {Allart}, \& {Bourrier}}]{Mansfield2018}
{Mansfield}, M., {Bean}, J.~L., {Oklop{\v{c}}i{\'c}}, A., {et~al.} 2018, \apjl,
  868, L34, \dodoi{10.3847/2041-8213/aaf166}

\bibitem[{{Mazeh} {et~al.}(2016){Mazeh}, {Holczer}, \& {Faigler}}]{Mazeh2016}
{Mazeh}, T., {Holczer}, T., \& {Faigler}, S. 2016, \aap, 589, A75,
  \dodoi{10.1051/0004-6361/201528065}

\bibitem[{{Mordasini}(2020)}]{Mordasini2020}
{Mordasini}, C. 2020, \aap, 638, A52, \dodoi{10.1051/0004-6361/201935541}

\bibitem[{{Murray-Clay} {et~al.}(2009){Murray-Clay}, {Chiang}, \&
  {Murray}}]{MClay2009}
{Murray-Clay}, R.~A., {Chiang}, E.~I., \& {Murray}, N. 2009, \apj, 693, 23,
  \dodoi{10.1088/0004-637X/693/1/23}

\bibitem[{{Nakayama} {et~al.}(2022){Nakayama}, {Ikoma}, \&
  {Terada}}]{Nakayama2022}
{Nakayama}, A., {Ikoma}, M., \& {Terada}, N. 2022, \apj, 937, 72,
  \dodoi{10.3847/1538-4357/ac86ca}

\bibitem[{{Nortmann} {et~al.}(2018){Nortmann}, {Pall{\'e}}, {Salz},
  {Sanz-Forcada}, {Nagel}, {Alonso-Floriano}, {Czesla}, {Yan}, {Chen},
  {Snellen}, {Zechmeister}, {Schmitt}, {L{\'o}pez-Puertas}, {Casasayas-Barris},
  {Bauer}, {Amado}, {Caballero}, {Dreizler}, {Henning}, {Lamp{\'o}n}, {Montes},
  {Molaverdikhani}, {Quirrenbach}, {Reiners}, {Ribas}, {S{\'a}nchez-L{\'o}pez},
  {Schneider}, \& {Zapatero Osorio}}]{Nortmann2018}
{Nortmann}, L., {Pall{\'e}}, E., {Salz}, M., {et~al.} 2018, Science, 362, 1388,
  \dodoi{10.1126/science.aat5348}

\bibitem[{{Oklop{\v{c}}i{\'c}}(2019)}]{Oklopcic2019}
{Oklop{\v{c}}i{\'c}}, A. 2019, \apj, 881, 133, \dodoi{10.3847/1538-4357/ab2f7f}

\bibitem[{{Oklop{\v{c}}i{\'c}} \& {Hirata}(2018)}]{Oklopcic2018}
{Oklop{\v{c}}i{\'c}}, A., \& {Hirata}, C.~M. 2018, \apjl, 855, L11,
  \dodoi{10.3847/2041-8213/aaada9}

\bibitem[{{Orell-Miquel} {et~al.}(2022){Orell-Miquel}, {Murgas}, {Pall{\'e}},
  {Lamp{\'o}n}, {L{\'o}pez-Puertas}, {Sanz-Forcada}, {Nagel}, {Kaminski},
  {Casasayas-Barris}, {Nortmann}, {Luque}, {Molaverdikhani}, {Sedaghati},
  {Caballero}, {Amado}, {Bergond}, {Czesla}, {Hatzes}, {Henning},
  {Khalafinejad}, {Montes}, {Morello}, {Quirrenbach}, {Reiners}, {Ribas},
  {S{\'a}nchez-L{\'o}pez}, {Schweitzer}, {Stangret}, {Yan}, \& {Zapatero
  Osorio}}]{Orell2022}
{Orell-Miquel}, J., {Murgas}, F., {Pall{\'e}}, E., {et~al.} 2022, \aap, 659,
  A55, \dodoi{10.1051/0004-6361/202142455}

\bibitem[{{Owen} \& {Wu}(2013)}]{Owen2013}
{Owen}, J.~E., \& {Wu}, Y. 2013, \apj, 775, 105,
  \dodoi{10.1088/0004-637X/775/2/105}

\bibitem[{{Owen} \& {Wu}(2017)}]{Owen2017}
---. 2017, \apj, 847, 29, \dodoi{10.3847/1538-4357/aa890a}

\bibitem[{{Palle} {et~al.}(2020){Palle}, {Nortmann}, {Casasayas-Barris},
  {Lamp{\'o}n}, {L{\'o}pez-Puertas}, {Caballero}, {Sanz-Forcada}, {Lara},
  {Nagel}, {Yan}, {Alonso-Floriano}, {Amado}, {Chen}, {Cifuentes},
  {Cort{\'e}s-Contreras}, {Czesla}, {Molaverdikhani}, {Montes}, {Passegger},
  {Quirrenbach}, {Reiners}, {Ribas}, {S{\'a}nchez-L{\'o}pez}, {Schweitzer},
  {Stangret}, {Zapatero Osorio}, \& {Zechmeister}}]{Palle2020}
{Palle}, E., {Nortmann}, L., {Casasayas-Barris}, N., {et~al.} 2020, \aap, 638,
  A61, \dodoi{10.1051/0004-6361/202037719}

\bibitem[{{Paragas} {et~al.}(2021){Paragas}, {Vissapragada}, {Knutson},
  {Oklop{\v{c}}i{\'c}}, {Chachan}, {Greklek-McKeon}, {Dai}, {Tinyanont}, \&
  {Vasisht}}]{Paragas2021}
{Paragas}, K., {Vissapragada}, S., {Knutson}, H.~A., {et~al.} 2021, \apjl, 909,
  L10, \dodoi{10.3847/2041-8213/abe706}

\bibitem[{{Parker}(1958)}]{Parker1958}
{Parker}, E.~N. 1958, \apj, 128, 664, \dodoi{10.1086/146579}

\bibitem[{{Poppenhaeger}(2022)}]{Poppenhaeger2022}
{Poppenhaeger}, K. 2022, \mnras, 512, 1751, \dodoi{10.1093/mnras/stac507}

\bibitem[{{Salz} {et~al.}(2016){Salz}, {Czesla}, {Schneider}, \&
  {Schmitt}}]{Salz2016}
{Salz}, M., {Czesla}, S., {Schneider}, P.~C., \& {Schmitt}, J.~H.~M.~M. 2016,
  \aap, 586, A75, \dodoi{10.1051/0004-6361/201526109}

\bibitem[{{Shaikhislamov} {et~al.}(2021){Shaikhislamov}, {Khodachenko},
  {Lammer}, {Berezutsky}, {Miroshnichenko}, \&
  {Rumenskikh}}]{Shaikhislamov2021}
{Shaikhislamov}, I.~F., {Khodachenko}, M.~L., {Lammer}, H., {et~al.} 2021,
  \mnras, 500, 1404, \dodoi{10.1093/mnras/staa2367}

\bibitem[{{Sing} {et~al.}(2019){Sing}, {Lavvas}, {Ballester}, {Lecavelier des
  Etangs}, {Marley}, {Nikolov}, {Ben-Jaffel}, {Bourrier}, {Buchhave}, {Deming},
  {Ehrenreich}, {Mikal-Evans}, {Kataria}, {Lewis}, {L{\'o}pez-Morales},
  {Garc{\'\i}a Mu{\~n}oz}, {Henry}, {Sanz-Forcada}, {Spake}, {Wakeford}, \&
  {PanCET Collaboration}}]{Sing2019}
{Sing}, D.~K., {Lavvas}, P., {Ballester}, G.~E., {et~al.} 2019, \aj, 158, 91,
  \dodoi{10.3847/1538-3881/ab2986}

\bibitem[{{Spake} {et~al.}(2021){Spake}, {Oklop{\v{c}}i{\'c}}, \&
  {Hillenbrand}}]{Spake2021}
{Spake}, J.~J., {Oklop{\v{c}}i{\'c}}, A., \& {Hillenbrand}, L.~A. 2021, \aj,
  162, 284, \dodoi{10.3847/1538-3881/ac178a}

\bibitem[{{Spake} {et~al.}(2018){Spake}, {Sing}, {Evans}, {Oklop{\v{c}}i{\'c}},
  {}, {Bourrier}, {Kreidberg}, {Rackham}, {Irwin}, {Ehrenreich}, {Wyttenbach},
  {Wakeford}, {Zhou}, {Chubb}, {Nikolov}, {Goyal}, {Henry}, {Williamson},
  {Blumenthal}, {Anderson}, {Hellier}, {Charbonneau}, {Udry}, \&
  {Madhusudhan}}]{Spake2018}
{Spake}, J.~J., {Sing}, D.~K., {Evans}, T.~M., {et~al.} 2018, \nat, 557, 68,
  \dodoi{10.1038/s41586-018-0067-5}

\bibitem[{{Spake} {et~al.}(2022){Spake}, {Oklop{\v{c}}i{\'c}}, {Hillenbrand},
  {Knutson}, {Kasper}, {Dai}, {Orell-Miquel}, {Vissapragada}, {Zhang}, \&
  {Bean}}]{Spake2022}
{Spake}, J.~J., {Oklop{\v{c}}i{\'c}}, A., {Hillenbrand}, L.~A., {et~al.} 2022,
  \apjl, 939, L11, \dodoi{10.3847/2041-8213/ac88c9}

\bibitem[{{Szab{\'o}} \& {Kiss}(2011)}]{Szabo2011}
{Szab{\'o}}, G.~M., \& {Kiss}, L.~L. 2011, \apj, 727, L44,
  \dodoi{10.1088/2041-8205/727/2/L44}

\bibitem[{{Vidal-Madjar} {et~al.}(2003){Vidal-Madjar}, {Lecavelier des Etangs},
  {D{\'e}sert}, {Ballester}, {Ferlet}, {H{\'e}brard}, \& {Mayor}}]{Vidal2003}
{Vidal-Madjar}, A., {Lecavelier des Etangs}, A., {D{\'e}sert}, J.-M., {et~al.}
  2003, \nat, 422, 143, \dodoi{10.1038/nature01448}

\bibitem[{{Vidal-Madjar} {et~al.}(2004){Vidal-Madjar}, {D{\'e}sert},
  {Lecavelier des Etangs}, {H{\'e}brard}, {Ballester}, {Ehrenreich}, {Ferlet},
  {McConnell}, {Mayor}, \& {Parkinson}}]{Vidal2004}
{Vidal-Madjar}, A., {D{\'e}sert}, J.-M., {Lecavelier des Etangs}, A., {et~al.}
  2004, \apjl, 604, L69, \dodoi{10.1086/383347}

\bibitem[{{Vidal-Madjar} {et~al.}(2013){Vidal-Madjar}, {Huitson}, {Bourrier},
  {D{\'e}sert}, {Ballester}, {Lecavelier des Etangs}, {Sing}, {Ehrenreich},
  {Ferlet}, {H{\'e}brard}, \& {McConnell}}]{Vidal2013}
{Vidal-Madjar}, A., {Huitson}, C.~M., {Bourrier}, V., {et~al.} 2013, \aap, 560,
  A54, \dodoi{10.1051/0004-6361/201322234}

\bibitem[{{Vidotto} \& {Cleary}(2020)}]{Vidotto2020}
{Vidotto}, A.~A., \& {Cleary}, A. 2020, \mnras, 494, 2417,
  \dodoi{10.1093/mnras/staa852}

\bibitem[{{Villarreal D'Angelo} {et~al.}(2021){Villarreal D'Angelo}, {Vidotto},
  {Esquivel}, {Hazra}, \& {Youngblood}}]{Villarreal2021}
{Villarreal D'Angelo}, C., {Vidotto}, A.~A., {Esquivel}, A., {Hazra}, G., \&
  {Youngblood}, A. 2021, \mnras, 501, 4383, \dodoi{10.1093/mnras/staa3867}

\bibitem[{Virtanen {et~al.}(2020)Virtanen, Gommers, Oliphant, Haberland, Reddy,
  Cournapeau, Burovski, Peterson, Weckesser, Bright, {van der Walt}, Brett,
  Wilson, Millman, Mayorov, Nelson, Jones, Kern, Larson, Carey, Polat, Feng,
  Moore, {VanderPlas}, Laxalde, Perktold, Cimrman, Henriksen, Quintero, Harris,
  Archibald, Ribeiro, Pedregosa, {van Mulbregt}, \& {SciPy 1.0
  Contributors}}]{2020SciPy-NMeth}
Virtanen, P., Gommers, R., Oliphant, T.~E., {et~al.} 2020, Nature Methods, 17,
  261, \dodoi{10.1038/s41592-019-0686-2}

\bibitem[{{Vissapragada} {et~al.}(2022{\natexlab{a}}){Vissapragada}, {Knutson},
  {dos Santos}, {Wang}, \& {Dai}}]{Vissapragada2022a}
{Vissapragada}, S., {Knutson}, H.~A., {dos Santos}, L.~A., {Wang}, L., \&
  {Dai}, F. 2022{\natexlab{a}}, \apj, 927, 96, \dodoi{10.3847/1538-4357/ac4e8a}

\bibitem[{{Vissapragada} {et~al.}(2021){Vissapragada}, {Stef{\'a}nsson},
  {Greklek-McKeon}, {Oklop{\v{c}}i{\'c}}, {Knutson}, {Ninan}, {Mahadevan},
  {Ca{\~n}as}, {Chachan}, {Cochran}, {Collins}, {Dai}, {David}, {Halverson},
  {Hawley}, {Hebb}, {Kanodia}, {Kowalski}, {Livingston}, {Maney}, {Metcalf},
  {Morley}, {Ramsey}, {Robertson}, {Roy}, {Spake}, {Schwab}, {Terrien},
  {Tinyanont}, {Vasisht}, \& {Wisniewski}}]{Vissapragada2021}
{Vissapragada}, S., {Stef{\'a}nsson}, G., {Greklek-McKeon}, M., {et~al.} 2021,
  \aj, 162, 222, \dodoi{10.3847/1538-3881/ac1bb0}

\bibitem[{{Vissapragada} {et~al.}(2022{\natexlab{b}}){Vissapragada}, {Knutson},
  {Greklek-McKeon}, {Oklop{\v{c}}i{\'c}}, {Dai}, {dos Santos}, {Jovanovic},
  {Mawet}, {Millar-Blanchaer}, {Paragas}, {Spake}, {Tinyanont}, \&
  {Vasisht}}]{Vissapragada2022b}
{Vissapragada}, S., {Knutson}, H.~A., {Greklek-McKeon}, M., {et~al.}
  2022{\natexlab{b}}, \aj, 164, 234, \dodoi{10.3847/1538-3881/ac92f2}

\bibitem[{{Watson} {et~al.}(1981){Watson}, {Donahue}, \& {Walker}}]{Watson1981}
{Watson}, A.~J., {Donahue}, T.~M., \& {Walker}, J.~C.~G. 1981, \icarus, 48,
  150, \dodoi{10.1016/0019-1035(81)90101-9}

\bibitem[{{Zhang} {et~al.}(2023){Zhang}, {Knutson}, {Dai}, {Wang}, {Ricker},
  {Schwarz}, {Mann}, \& {Collins}}]{Zhang2023}
{Zhang}, M., {Knutson}, H.~A., {Dai}, F., {et~al.} 2023, \aj, 165, 62,
  \dodoi{10.3847/1538-3881/aca75b}

\bibitem[{{Zhang} {et~al.}(2022){Zhang}, {Knutson}, {Wang}, {Dai}, {dos
  Santos}, {Fossati}, {Henry}, {Ehrenreich}, {Alibert}, {Hoyer}, {Wilson}, \&
  {Bonfanti}}]{Zhang2022}
{Zhang}, M., {Knutson}, H.~A., {Wang}, L., {et~al.} 2022, \aj, 163, 68,
  \dodoi{10.3847/1538-3881/ac3f3b}

\end{thebibliography}
\bibliographystyle{aasjournal}



\end{document}